\documentclass[fleqn,usenatbib]{mnras}

\usepackage{newtxtext,newtxmath}

\usepackage[T1]{fontenc}

\DeclareRobustCommand{\VAN}[3]{#2}
\let\VANthebibliography\thebibliography
\def\thebibliography{\DeclareRobustCommand{\VAN}[3]{##3}\VANthebibliography}

\usepackage{graphicx}	
\usepackage{amsmath}	
\usepackage{threeparttable}
\usepackage{diagbox}

\newcommand{\meiso}{E_{\gamma{\rm , iso}}}
\newcommand{\thetaobs}{$\theta_{\rm obs}$}
\newcommand{\mthetaobs}{\theta_{\rm obs}}
\newcommand{\thetainc}{$\theta_{\rm inc}$}
\newcommand{\mthetainc}{\theta_{\rm inc}}
\newcommand{\msol}{$M_{\odot}$}
\newcommand{\degsq}{deg$^2$}
\defcitealias{Evans16a}{Paper I}
\defcitealias{Evans16b}{Paper II}
\defcitealias{Nicholl21}{N21}

\title[Panning for gold with \textit{Swift}]{Panning for gold with the \textit{Neil Gehrels Swift Observatory}: an optimal strategy for finding the counterparts to gravitational wave events}

\author[R. A. J. Eyles-Ferris et al.]{
R. A. J. Eyles-Ferris,$^{1}$\thanks{E-mail: raje1@leicester.ac.uk} P. A. Evans,$^{1}$ A. A. Breeveld,$^{2}$ S. B. Cenko,$^{3,4}$ S. Dichiara,$^{5}$, J. A. Kennea,$^{5}$
\and N. J. Klingler,$^{6,3,7}$ N. P. M. Kuin,$^{2}$ F. E. Marshall,$^{3}$ S. R. Oates,$^{8}$ M. J. Page,$^{2}$ G. Raman,$^{5}$ S. Ronchini,$^{5,9}$
\and M. H. Siegel,$^{5}$ A. Tohuvavohu,$^{10,11,12}$ S. Campana,$^{13}$ V. D'Elia,$^{14,15}$ D. H. Hartmann,$^{16}$ J. P. Osborne,$^{1}$
\and K. L. Page,$^{1}$ M. De Pasquale,$^{17}$ and E. Troja$^{18, 19}$
\\
$^{1}$School of Physics and Astronomy, University of Leicester, University Road, Leicester, LE1 7RH, UK\\
$^{2}$Mullard Space Science Laboratory, University College London, Holmbury St Mary, Dorking, Surrey, RH5 6NT, UK\\
$^{3}$Astrophysics Science Division, NASA Goddard Space Flight Center, Greenbelt, MD 20771, USA\\
$^{4}$Joint Space-Science Institute, Computer and Space Sciences Building, University of Maryland, College Park, MD 20742, USA\\
$^{5}$Department of Astronomy and Astrophysics, The Pennsylvania State University, University Park, PA 16802, USA\\
$^{6}$Center for Space Sciences and Technology, University of Maryland, Baltimore County, Baltimore, MD, 21250, USA\\
$^{7}$Center for Research and Exploration in Space Science and Technology, NASA Goddard Space Flight Center, Greenbelt, Maryland 20771, USA\\
$^{8}$Physics Department, Lancaster University, Bailrigg, Lancaster, LA1 4YB, UK\\
$^{9}$Institute for Gravitation \& the Cosmos, The Pennsylvania State University, University Park, PA 16802, USA\\
$^{10}$David A. Dunlap Department of Astronomy and Astrophysics, University of Toronto, Toronto M5S 3H7, Canada\\
$^{11}$Dunlap Institute for Astronomy and Astrophysics, University of Toronto, Toronto M5S 3H7, Canada\\
$^{12}$Cahill Center for Astronomy and Astrophysics, California Institute of Technology, Pasadena, CA 91125, USA\\
$^{13}$INAF - Osservatorio Astronomico di Brera, Via Bianchi 46, I-23807 Merate (LC), Italy\\
$^{14}$Space Science Data Center (SSDC) - Agenzia Spaziale Italiana (ASI), I-00133 Roma, Italy\\
$^{15}$INAF - Osservatorio Astronomico di Roma, via Frascati 33, 00040 Monte Porzio Catone, Italy\\
$^{16}$Department of Physics and Astronomy, Clemson University, Clemson, SC 29634-0978, USA\\
$^{17}$MIFT Department, University of Messina, Vi F. S. D'Alcontres 31, 98166, Italy\\
$^{18}$Dipartimento di Fisica, Universit\`a Degli Studi di Roma - Tor Vergata, via della Ricera Scientifica 1, 00100 Rome, Italy\\
$^{18}$INAF - National Institute of Astrophysics, Rome, Italy\\
}

\date{Accepted XXX. Received YYY; in original form ZZZ}

\pubyear{2024}

\begin{document}
\label{firstpage}
\pagerange{\pageref{firstpage}--\pageref{lastpage}}
\maketitle

\begin{abstract}
The LIGO, Virgo and KAGRA gravitational wave observatories are currently undertaking their O4 observing run offering the opportunity to discover new electromagnetic counterparts to gravitational wave events. We examine the capability of the \textit{Neil Gehrels Swift Observatory} (\textit{Swift}) to respond to these triggers, primarily binary neutron star mergers, with both the UV/Optical Telescope (UVOT) and the X-ray Telescope (XRT). We simulate \textit{Swift}'s response to a trigger under different strategies using model skymaps, convolving these with the 2MPZ catalogue to produce an ordered list of observing fields, deriving the time taken for \textit{Swift} to reach the correct field and simulating the instrumental responses to modelled kilonovae and short gamma-ray burst afterglows.  We find that UVOT, using the \textit{u} filter with an exposure time of order 120 s, is optimal for most follow-up observations and that we are likely to detect counterparts in $\sim6$\% of all binary neutron star triggers detectable by LVK in O4. We find that the gravitational wave 90\% error area and measured distance to the trigger allow us to select optimal triggers to follow-up. Focussing on sources less than 300 Mpc away, or 500 Mpc if the error area is less than a few hundred square degrees, distances greater than previously assumed, offer the best opportunity for discovery by \textit{Swift} with $\sim5 - 30$\% of triggers having detection probabilities $\geq 0.5$. At even greater distances, we can further optimise our follow-up by adopting a longer 250 s or 500 s exposure time.
\end{abstract}




\section{Introduction}

GW 170817 was a true watershed moment in astronomy, the first time a gravitational wave (GW) detection was found to have an electromagnetic (EM) counterpart \citep[e.g.][]{Abbott17,Covino17,Cowperthwaite17,Evans17,Goldstein17,Kilpatrick17,Savchenko17,Smartt17,Tanvir17,Troja17}. However, at time of writing, it remains a unique event. The Laser Interferometer Gravitational-Wave Observatory (LIGO), the Virgo interferometer and Kamioka Gravitational Wave Detector (KAGRA), together referred to in this work as LVK, commenced their O4 run on 24 May 2023 offering the opportunity to finally make GW 170817 commonplace rather than unique. It is therefore crucial to devise an optimal strategy to follow-up GW detections and quickly discover any EM counterparts. Here we build on the work of \citet[][hereafter \citetalias{Evans16a}]{Evans16a} and \citet[][hereafter \citetalias{Evans16b}, see also \citealt{Evans16b_erratum}]{Evans16b}, which discuss the strategy for previous LVK runs. However, both works assumed that the afterglow from a short GRB would be the signal most likely to be detected by \textit{Swift} and therefore concentrated on \textit{Swift}'s X-Ray Telescope (XRT) as the discovery instrument. However, the afterglow of GW 170817 was not detected by the XRT\footnote{Although we note that \textit{Swift} didn't observe its position until $\sim14$ hours after the trigger.} but did have a UV bright kilonova \citep{Evans17} detected by the UV/Optical Telescope (UVOT).

A short GRB and a kilonova are the two EM transients that are typically expected to accompany the merger of a binary neutron star (BNS) or neutron star-black hole (NSBH) system, in addition to the gravitational waves detectable by LVK. Short GRBs result in two signals, the prompt gamma-ray emission that emerges from within a relativistic jet (the burst itself) and a synchrotron powered broadband afterglow resulting from jet interactions with the circumburst medium. However due to Doppler beaming, both these components are, at least initially, strongly directional, particularly at the wavelengths accessible to \textit{Swift}. While structured jets or cocoon emission \citep{Nakar17,Piro18} do result in more isotropic emission, the relatively narrow opening angle of the jet in a short GRB \citep[$\sim 15 \degr$,][]{Fong15} therefore means that only a very small fraction of the merger population detectable by LVK through the far more isotropic gravitational wave emission will have bright and detectable afterglows. Kilonovae, on the other hand, are powered by \textit{r}-process nucleosynthesis in the merger ejecta and emit much more isotropically \citep[e.g.][]{Metzger19}. While there is still significant viewing angle dependence due to the nature of the ejecta, a search focussed on kilonova emission might therefore be most effective. In this work, therefore, we focus primarily on the UVOT but also apply our analysis to the XRT.

In this paper, we examine the UVOT's capability in detecting a counterpart from a GW trigger during O4. We concentrate primarily on BNS mergers as their resultant kilonovae are significantly brighter and easier to detect than those expected from NSBH mergers \citep[e.g.][]{Gompertz23}, although we also briefly discuss these sources. Although we generally find the probability of a significant detection relatively low, we can improve our chances by using specific criteria such as the 90\% error area and the luminosity distance. This was already used to inform \textit{Swift}'s strategy and this work has aided in updating these criteria and the strategy accordingly. Selecting events using these criteria significantly enhances the chance of UVOT promptly discovering an EM counterpart and ensures \textit{Swift} remains effective across all aspects of its mission.

In Section \ref{sec:trigger_sims}, we detail the setup of our trigger simulations and their underlying data. We also describe \textit{Swift}'s response to a trigger. Section \ref{sec:lc_modelling} presents our modelling methodology for the afterglow and kilonova components of the light curve. We evaluate \textit{Swift}'s expected performance and how to optimise follow-up in Section \ref{sec:obs_strategy}. We discuss additional factors that could impact follow-up in Section \ref{sec:discussion} and present our conclusions in Section \ref{sec:conclusions}. Unless otherwise mentioned, we adopt a cosmology with $H_0 = 71$ km\,s$^{-1}$\,Mpc$^{-1}$, $\Omega_m = 0.27$ and $\Omega_\Lambda = 0.73$ throughout this paper.

\section{Simulated triggers}
\label{sec:trigger_sims}

To evaluate and optimise \textit{Swift}'s observing plan, we simulated 3688 triggers. To summarise, from simulations of observed GW events and their associated skymaps, we seed the event in a host galaxy or random point across the sky. The skymap is then shifted and convolved with known galaxies to produce an ordered list of fields to be observed. Our full procedure is detailed in the rest of this section.

\subsection{Gravitational wave simulations}

To examine BNS mergers, we used the dataset of \citet{Coughlin22}, which builds on the work and simulations of \citet{Singer14} and \citet{Singer16a}. This consists of 8,258 simulations of compact binary mergers, including both neutron stars (NSs) and black holes (BHs). Each of these simulations has an associated skymap derived using a range of instrument combinations based on the predicted duty cycle. The assumed sensitivity of each instrument is based on the predictions for O4. To limit our sample to mergers most capable of producing kilonovae, i.e. BNSs, we selected the simulations where the mass of each component was less than 3\msol, a total of 922 simulations. Each simulation was used to simulate four triggers hence our 3688 total triggers.

\begin{figure}
    \centering
    \includegraphics[width=\columnwidth]{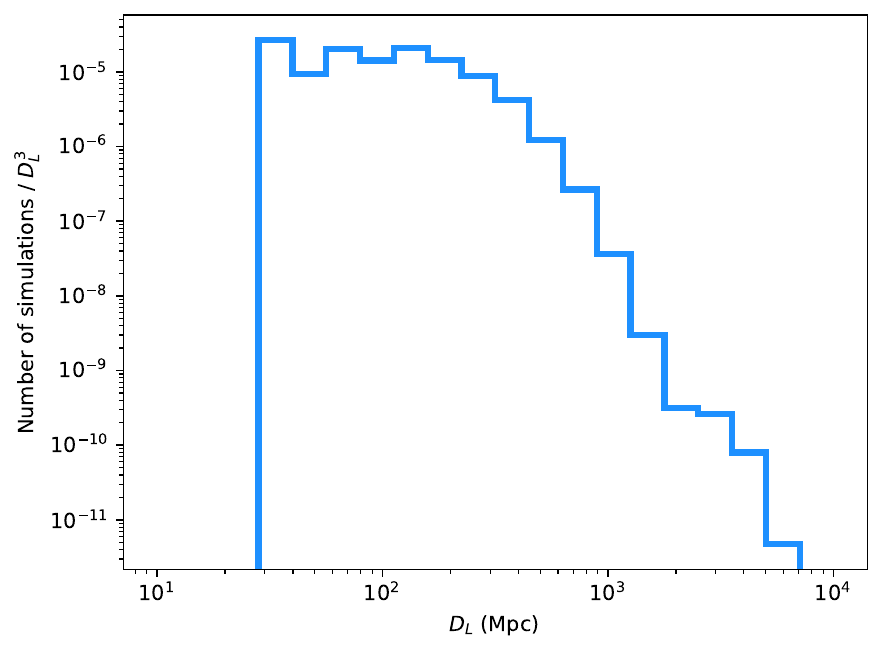}
    \caption{The distribution of luminosity distance for the simulation sample, normalised with the luminosity distance cubed.}
    \label{fig:sims_distances}
\end{figure}

We checked the properties of our sample of simulations and confirmed that, like the overall population, the mergers were randomly placed across the sky with an isotropic distribution. We found the mergers to be distributed uniformly in distance cubed out to $\sim350$ Mpc with a rapid decline at higher luminosity distances as shown in Figure \ref{fig:sims_distances}. The sensitivities of the observatories were expected to increase to 160 -- 190 Mpc for LIGO-Hanford and Livingston and 80 -- 115 Mpc for Virgo for the merger of two 1.4 \msol~NSs between O3 and O4. The simulation sample therefore covers a suitable range with higher distances attributable to the larger NS masses allowed. The inclinations of the systems have a bimodal distribution with peaks at $\sim30\degr$ and $\sim150\degr$. This is due to the dependency of the GW signal on inclination and is consistent with the distribution expected due to the Malmquist bias \citep[e.g.][]{Schutz11}.

\begin{figure*}
    \centering
    \includegraphics[width=\columnwidth]{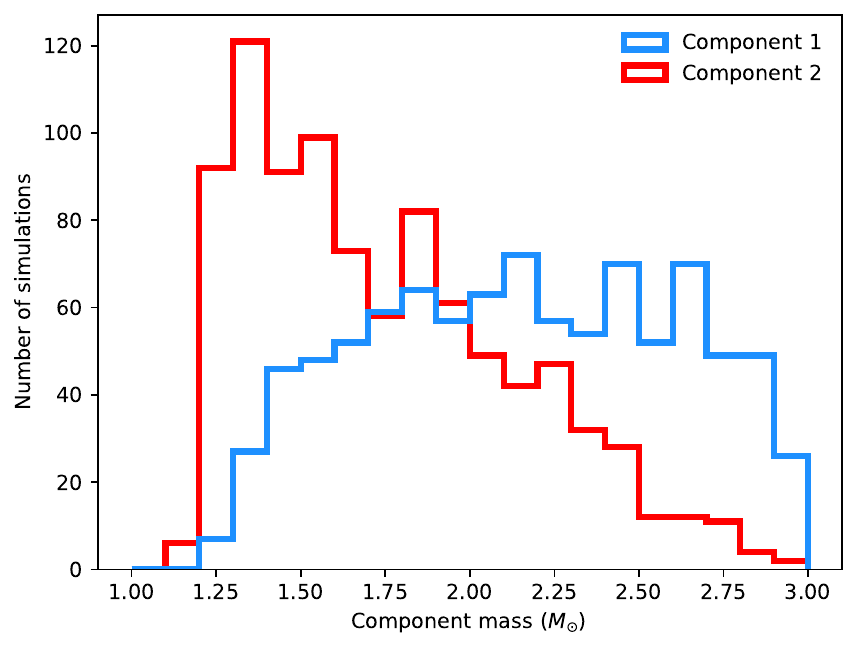}
    \includegraphics[width=\columnwidth]{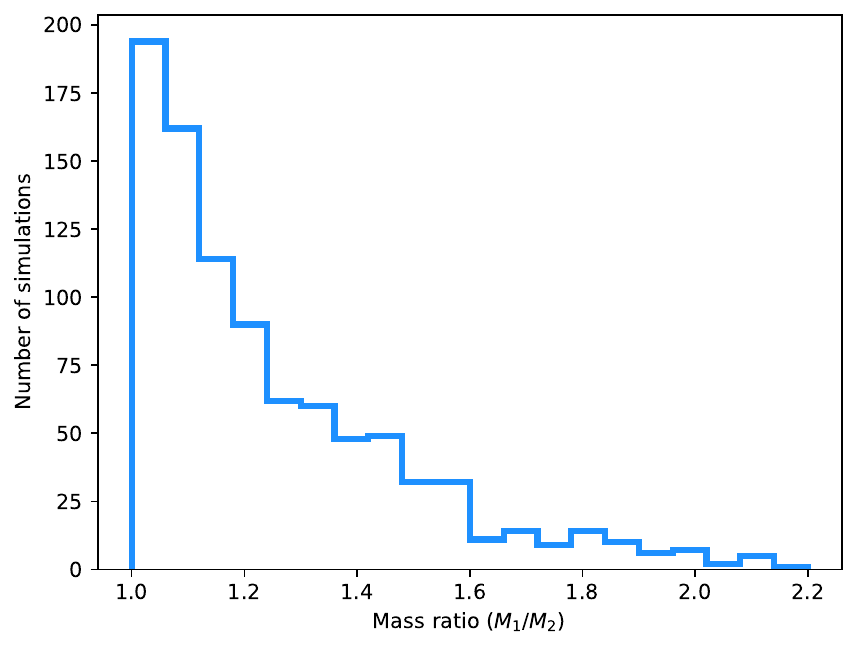}
    \caption{The distributions of component masses (left) and mass ratio (right) for the simulation sample. Note that component 1 is, by definition, the more massive of the two.}
    \label{fig:sims_masses}
\end{figure*}

The distributions of the masses of each component and their ratios are shown in Figure \ref{fig:sims_masses}. Component 1 is always more massive by definition and this leads to its relatively flat distribution and the low peak of component 2's distribution. Observationally, BNS systems within the Milky Way have been found to have mass ratios up to $\sim1.3$ \citep[e.g.][]{Tauris17,Ferdman20} and theory predicts ratios up to $\sim1.7$ \citep[e.g.][]{Wagg22}. The mass ratio distribution of the simulated triggers is somewhat broader than this. However, greater mass ratios may still be physically viable and they represent only a small minority within our sample. The masses and their ratios therefore cover a wide range of BNSs with diverse properties. 

Overall, our sample is consistent with expectations for the BNS population and is ideal for exploring the possibilities of such mergers across O4.

\subsection{Host selection}

A real BNS merger will, rather than be positioned randomly in space, be associated with a host galaxy which may be known or unknown. We therefore used the Two Micron All Sky Survey Photometric Redshift catalogue \citep[2MPZ,][]{Bilicki14} to determine if the source is placed in a catalogued galaxy and if it is, to select a suitable host for each simulated trigger to combine with its skymap. The completeness of the galaxy catalogue must therefore be taken into account when seeding a trigger into a galaxy. For 2MPZ, the completeness was derived in \citetalias{Evans16b} following \citet{Gehrels16}, by comparing the luminosity observed in the catalogue to that predicted and we summarise their method here.

The predicted integrated luminosity per unit volume of the local universe is described well by the integral of a Schechter function \citep{Schechter76},
\begin{equation}
\label{eq:schechter}
    \frac{d\mathcal{L}}{dV} = \int^{\infty}_0 \phi^* L^* x^{\alpha+1} e^{-x} dx
\end{equation}
where $x=L/L^*$, $\mathcal{L}$ is the total integrated luminosity across the universe, $L^*$ is the characteristic galaxy luminosity at which the luminosity function steepens significantly, $\phi^*$ is a number density and $\alpha$ describes the slope for $L < L^*$. These last three parameters are all measurable from observations. For the IR selected 2MPZ, the absolute \textit{K} band magnitudes ($M^*$ for a galaxy with luminosity $L^*$) are used to derive these parameters: $M^*_K = -23.39+5\log h$, $\alpha_K = -1.09$ and $\phi^*_K = 0.0116h^3$ \citep{Kochanek01}, where we have assumed $h = H_0 / 100 = 0.7$. Equation \ref{eq:schechter} can be further manipulated to derive the total luminosity within volume $V$ as
\begin{equation}
\label{eq:x_in_V}
    \frac{\mathcal{L}}{L^*} = \int^{\infty}_0 \phi^* x^{\alpha+1} e^{-x} dx V = \phi^* \Gamma(\alpha+2,0)V
\end{equation}
where $\Gamma$ is the incomplete gamma function\footnote{We note that this is only valid for $\alpha>-2$ and diverges for smaller values \citep{Gehrels16}.}.

Rather than the total IR magnitudes, 2MPZ contains the magnitude measured to the 20 mag arcsec$^{-2}$ isophote. We therefore corrected the catalogued values by $\Delta M_K=-0.2$ in order to match the total magnitude \citep{Kochanek01, Bilicki11}. The Galactic extinction observed for each source must also be accounted for. At higher extinctions, there is a natural bias towards redder sources as bluer galaxies are more likely to be obscured. The colours and therefore photometric redshifts are accordingly less reliable. For 2MPZ, this results in the photometric redshift, $z_{\rm phot}$, being underestimated by 0.004 compared to the spectroscopic redshift\footnote{In our chosen cosmology, this increases the luminosity distances by $\sim 18$ Mpc.}, $z_{\rm spec}$ for sources with $0.5 \leq E_{B - V} < 1.0$. At even higher extinctions, this offset becomes greater and the redshift calibration becomes very unreliable. To alleviate these issues, the 9638 (1.0\% of the catalogue) sources with $E_{B-V} > 1.0$ were cut from the sample. This also had the effect of eliminating sources affected by high stellar densities, which introduces an additional colour bias and affect the UVOT's UV filters more significantly than optical wavelengths. This final sample was used to derive the completeness of 2MPZ with distance, $C$, shown in Figure 5 of \citetalias{Evans16b}. We note that \citetalias{Evans16b} did identify that a small discrepancy between $z_{\rm phot}$ and $z_{\rm spec}$, $\sigma_{z_{\rm photo}} = z_{\rm spec} - z_{\rm photo}$, remained in this final sample, characterised as a broken power law,
\begin{equation}
\label{eq:photoz_error}
    \sigma_{z_{\rm photo}} = \begin{cases} 0.043\, z_{\rm photo}^{0.402} & z_{\rm photo}<0.10 \\ 0.023\, z_{\rm photo}^{0.140} & z_{\rm photo}\geq0.10 \end{cases}
\end{equation}
However, because it is more likely that the photometric redshifts are underestimates rather than overestimates, the completeness is actually greater than calculated. 

For a given trigger, therefore, the probability of it being in a known galaxy is simply the completeness at that distance i.e. $P_{\rm gal} = C$. If probabilistically determined to be located in a known galaxy\footnote{Due to high natal kick velocities, a trigger could occur a significant distance `outside' its host galaxy \citep[e.g.][]{Mandhai22}. However, at the luminosity distances involved, this offset is relatively insignificant and can therefore be neglected in our analysis.}, a catalogued galaxy is randomly selected to be the host weighted by the \textit{K} band luminosity as a proxy for the galaxies' stellar masses. We also require the host to be at an appropriate distance and a similar declination to the original position of the trigger. The original skymap is then shifted across the sky to match the trigger's new location (as the real position of the source is known relative to the skymap). By maintaining a similar declination, we minimise the effect of the shift on the GW localisation due to the locations of the GW detectors. This means we effectively only change the (fictional) time that a given trigger was detected which has no impact on the follow-up strategy. For sources that weren't selected to be in a host galaxy, their position and associated skymap were left unchanged from the original simulation.

\subsection{Simulated follow-up}
\label{sec:simulated_followup}

Having derived a trigger, we also model \textit{Swift's} planned response. This is again detailed in \citetalias{Evans16b} and its corresponding erratum \citep{Evans16b_erratum}, and we give an overview of it here. We note that this response is modelled as if it were a real trigger e.g. we do not know if the trigger is in a known host galaxy.

The GW skymaps are given in \textsc{healpix} format and each individual pixel, $p$, in the skymap has a probability of the trigger occuring within it, $P_{{\rm GW},p}$. Because a given trigger will be associated with a host galaxy this skymap can be convolved with a galaxy catalogue to more accurately capture the relative probability of each pixel. However, this means the completeness of the chosen galaxy catalogue also has a significant effect on the follow-up strategy as the trigger could be in an uncatalogued or a catalogued galaxy\footnote{There is also a possibility of chance alignment between a galaxy and the trigger. However, the chance is extremely low, particularly due to the fact that the distance is also constrained, and thus this effect is likely to be negligible and therefore ignored in our analysis.}. This means the effective probability for a given pixel is
\begin{equation}
\label{eq:pixel_probability}
    \mathcal{P}_{{\rm GW},p} = \mathcal{P}_{{\rm nogal},p} + \mathcal{P}_{{\rm gal},p}
\end{equation}
where $\mathcal{P}_{{\rm nogal},p}$ and $\mathcal{P}_{{\rm gal},p}$ are the probabilities of the event occurring in an uncatalogued and a catalogued galaxy within $p$ respectively.

Each pixel in the skymap has its own probability distribution of the distance to the trigger, $D$. Consequently, the completeness of each pixel is
\begin{equation}
\label{eq:pixel_completeness}
    C_p = \frac{\int P_p(D)C(D)dD}{\int P_p(D)dD}
\end{equation}
where $P_p(D)$ is the probability distribution of distance in $p$. The probability of the GW event occuring in an uncatalogued galaxy within a pixel is therefore simply
\begin{equation}
\label{eq:uncatalogued_in_pixel}
    \mathcal{P}_{{\rm nogal},p} = P_{{\rm GW},p} (1 - C_p)
\end{equation}
The probability of the trigger occurring in a catalogued galaxy in pixel $p$ is slightly more complex, however. Essentially, $\mathcal{P}_{{\rm gal},p}$ is the probability of the pixel in the skymap multiplied by the fraction of total galaxy luminosity within the volume defined by the pixel and its distance distribution. It is given by
\begin{equation}
\label{eq:catalogued_in_pixel}
    \mathcal{P}_{{\rm gal},p} = P_{{\rm GW},p} C_p N \sum_g \left( \mathcal{P}(g|P_p(D)) \frac{L_g}{L_{\rm tot}}\right)
\end{equation}
where the sum is over all galaxies in the pixel, $L_g$ is the luminosity of the galaxy divided by the number of pixels it covers, $N$ is a normalisation factor such that $\sum_p \mathcal{P}_{{\rm gal},p} = C$, and $L_{\rm tot}$ is the total catalogued galaxy luminosity within the volume defined by the skymap and distance distribution. This means that $\sum_g \frac{L_g}{L_{\rm tot}}$ is effectively the relative probability of the galaxies in a specific pixel compared to any other pixel's galaxies. $N$ and $L_{\rm tot}$ are given by
\begin{equation}
\label{eq:N}
    N= \frac{\sum_p P_{{\rm GW},p} C_p}{\sum_p \left( P_{{\rm GW},p} C_p \sum_g \left(\mathcal{P}(g|P_p(D))\frac{L_g}{L_{\rm tot}}\right)\right)}
\end{equation}
and
\begin{equation}
\label{eq:Ltot}
    L_{\rm tot} = \sum_p \sum_g \mathcal{P}(g|P_p(D))L_g
\end{equation}
where $\mathcal{P}(g|P_p(D)) = \int P_p(D)P_g(D)dD$ is the probability that a given galaxy $g$ resides at an appropriate distance and $P_g(D)$ is the probability distribution of $g$'s distance. Because 2MPZ does not include errors on its redshifts, we assumed these distributions to be Gaussians. For sources with photometric redshifts, the error was taken from Equation \ref{eq:photoz_error} while for spectroscopic redshifts, the peculiar velocity was assumed to be the dominant source of error. Assuming a characteristic velocity of 500 km s$^{-1}$ gives a $\sigma$ of $500 / H_0 = 7.4$ Mpc (where $H_0=70$ km s$^{-1}$ Mpc$^{-1}$ has been assumed) to assign to the Gaussian.

\begin{figure}
    \centering
    \includegraphics[width=\columnwidth]{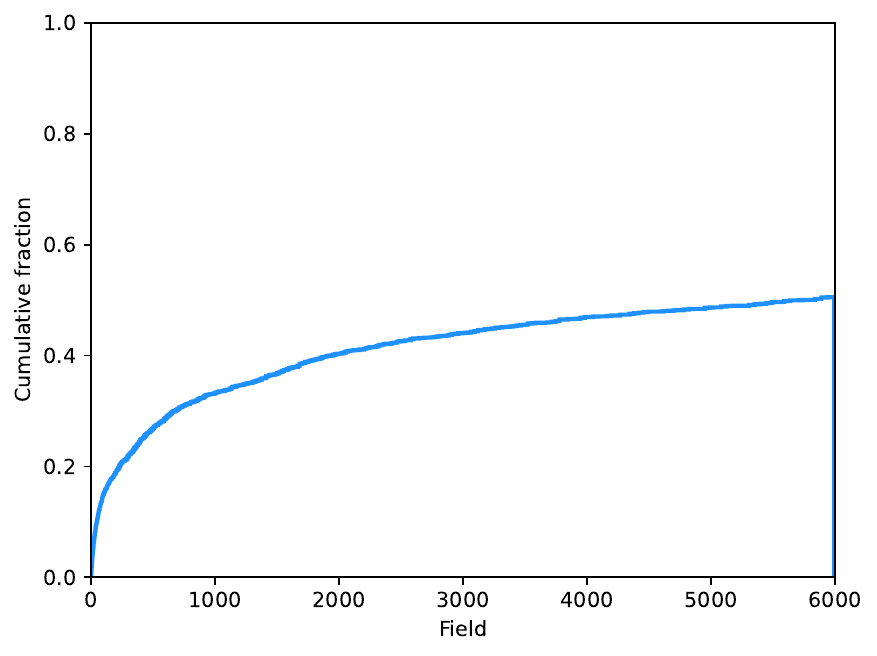}
    \caption{The cumulative distribution of simulated triggers against the field in which the source is contained.}
    \label{fig:fieldno}
\end{figure}

The convolution of the skymap and 2MPZ results in a new probability map. Following \citetalias{Evans16a}, this map can be divided into fields, containing a large number of pixels from the convolved map, that \textit{Swift} would observe in probability order. We can then model \textit{Swift}'s progress over these fields to determine the time at which it will reach the field containing the trigger's position, $t_{\rm reach}$. We simulate only to the 6000th field - if \textit{Swift} has not yet reached the transient, it can safely be assumed it would not be discovered as this will be several days after the trigger and any transient fades very rapidly. In Figure \ref{fig:fieldno}, we plot the distribution of the field which contains the source finding that 33.2\% of triggers lie within the first 1000 fields while 50.5\% of triggers are reached within the full 6000 fields simulated. This is somewhat lower than expected as in theory we use the same parameters to order galaxies when both seeding and searching. However, to ensure the skymaps are realistic relative to the positions of the LVK detectors, the seed galaxy is on roughly the same declination as the trigger in the original sky map. This means the seed galaxy is selected from a subset of galaxies rather than the much larger full population that is searched over and could introduce a small systematic bias that accounts for the proportion of triggers reached. It is also possible that flattening in the mass function of 2MPZ galaxies means that this is actually a purely stochastic effect, however, it is difficult to quantify which effect is more significant and whether one dominates over the other. We also ignore triggering criteria at this point - including these will significantly improve these fractions as discussed below.

To make our simulations as realistic as possible, we selected trigger times for each event and calculated the visibility windows of each field. Using realistic models of \textit{Swift}'s slewing capability we then created an optimised observing plan, thus obtaining the times at which each field would be observed in a real-world scenario, $t_{\rm reach}$.

There are three components to $t_{\rm reach}$: the time for \textit{Swift} to start observing, the cumulative time for it to slew between each field, and the cumulative exposure time to observe each field. Previously, in \citetalias{Evans16a}, it was assumed that the time for \textit{Swift} to start observing, $t_{\rm start}$, was best drawn from a Lorentz distribution based on the follow-up of IceCube neutrino triggers \citep{Evans15}. This distribution was indeed similar to the real values during O2 \citep{Klingler19} and O3 \citep{Page20, Oates21}. This delay primarily came from the time for the observing command upload to arrive at \textit{Swift}. However, the \textit{Swift} Mission Operations Centre have developed a system of `continuous commanding' over the Tracking and Data Relay Satellite System (TDRSS), removing the need for a ground-station pass and dramatically reducing the time between the trigger and the observing command upload \citep{Tohuvavohu24}. We therefore drew $t_{\rm start}$ from a normal distribution,
\begin{equation}
\label{eq:t_start}
    t_{\rm start} \sim \mathcal{N}(3600,1800)\,{\rm s}
\end{equation}
with a lower limit of 1200 s to account for the processing of the skymap and construction of the list of fields.

The time for \textit{Swift} to slew has also changed since the previous LVK runs. Following a reaction wheel failure in early 2022 \citep{Cenko22}, the response speed of \textit{Swift} is reduced and we therefore approximate the time to slew between two fields as 
\begin{equation}
\label{eq:t_slew}
    t_{\rm slew} = 25 + 2\delta \,{\rm s}
\end{equation}
where $\delta$ is the angular separation in degrees between the two fields.

Finally, the exposure time, $t_{\rm exp}$, for each field is controlled as part of the follow-up strategy. The exposure times and how they should vary in time to optimise \textit{Swift}'s follow-up are examined in Section \ref{sec:exposure_times}.

For a given trigger and the field it is contained within, the time to reach it is therefore
\begin{equation}
\label{eq:t_reach}
    t_{\rm reach} = t_{\rm start} + \sum_s t_{\rm slew} + \sum_e t_{\rm exp}
\end{equation}
summing over all slews, $s$, and exposures, $e$. We do note that our follow-up is likely to be interrupted to investigate candidates reported by other teams and $t_{\rm reach}$ increased accordingly. However, this is impossible to quantify and we therefore ignore this factor in the following analysis.

\section{Light curve modelling}
\label{sec:lc_modelling}

Having determined when \textit{Swift} will reach and observe the trigger, we must next evaluate what the electromagnetic transient will look like at that time. We assume there to be two components to this transient, a kilonova and an afterglow from an accompanying short GRB. We model these components independently and combine them to produce a final combined count rate for the transient. For simplicity, we assume the centre time of the observation to be representative of the transient's behaviour through the observation.

\subsection{Kilonova component}
\label{sec:kilonova_model}

\begin{figure}
    \centering
    \includegraphics[width=\columnwidth]{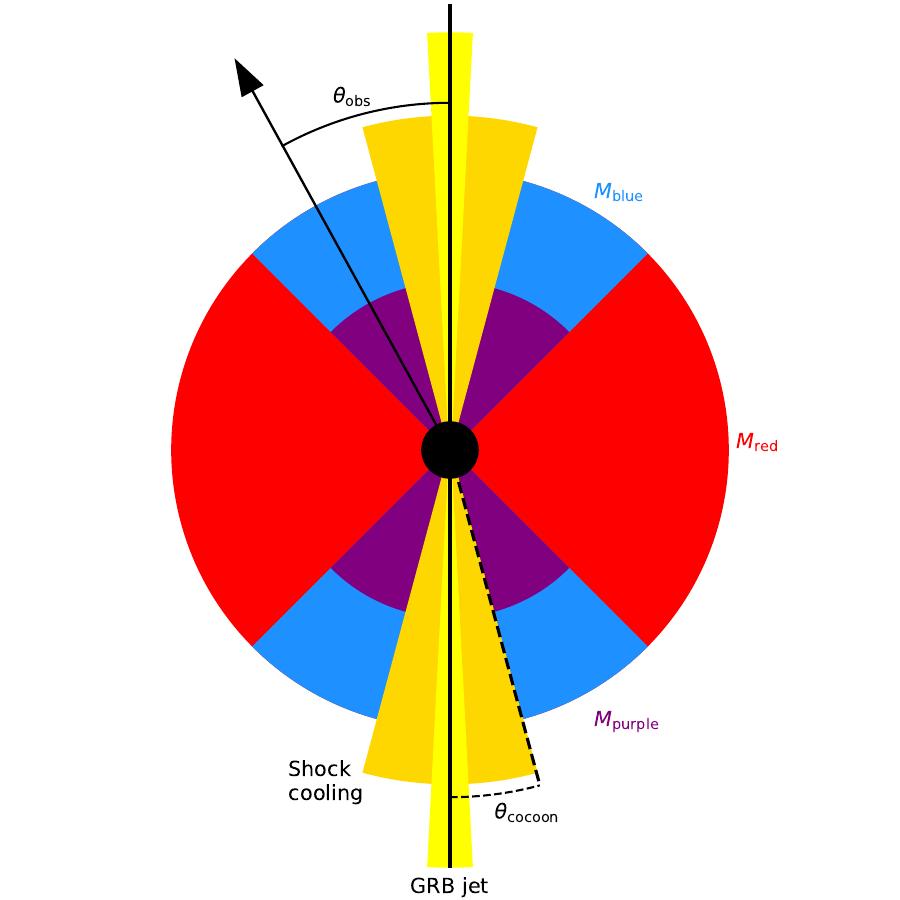}
    \caption{A sketch of the kilonova model assumed in \citetalias{Nicholl21}. The colours indicate the viewing angles of the various components and the black point represents the merger remnant.}
    \label{fig:kn_sketch}
\end{figure}

Their tiny observed population mean that kilonovae remain poorly understood and models of their emission are diverse in behaviour \citep[e.g.][]{Rossi22,Troja23}. For instance, toy models with assumed ejecta velocity distributions and opacities can approximate much of the behaviour of a kilonova \citep{Metzger19}. More advanced radiation transfer models, on the other hand, can fully predict the angular dependence of luminosity and spectra \citep{Kasen17,Bulla19,Korobkin21,Wollaeger21,Bulla23}. Arguably the most successful models, however, combine inferences and predictions from both the GW and EM data \citep{Coughlin19,Dietrich20,Breschi21}. Here we use the analytical BNS models of \citet[][hereafter \citetalias{Nicholl21}]{Nicholl21} which offer a distinct advantage over these options - they are designed to generate a light curve from only parameters inferred from a GW signal. These models are therefore ideal for use in our planning and during O4 itself.

The models are implemented in the \textsc{MOSFiT v.1.1.8} package \citep{Guillochon18}, specifically as the \texttt{bns\_generative} model, building on and utilising previous \textsc{MOSFiT} models of kilonovae \citep{Cowperthwaite17,Villar17}. The merger of the neutron stars produces ejecta components primarily consisting of dynamical ejecta and post-merger ejecta. From its primary inputs of the redshift, luminosity distance, $D_L$, chirp mass, $M_c$, mass ratio, $q$, and observation angle, \thetaobs, the model analytically predicts the masses, velocities and opacities of these components. The models of \citet{Cowperthwaite17} and \citet{Villar17} are then used to predict the luminosities of each component. We show a sketch of the modelled kilonova in Figure \ref{fig:kn_sketch} (see \citetalias{Nicholl21} for a significantly more detailed version). For this work, the parameters above were taken from the GW simulations, deriving \thetaobs~from the binary inclination angle, \thetainc. As the kilonova model assumes vertical symmetry, for $\mthetainc<90\degr$, $\mthetaobs=\mthetainc$, and for $\mthetainc>90\degr$, $\mthetaobs=180\degr-\mthetainc$.

The dynamical ejecta result from two interactions: the extreme tidal forces overcoming the NSs' self gravity as they orbit each other and material ejected by shocks at the contact interface. \citetalias{Nicholl21} use the prescription of \citet{Dietrich17b} for the total mass of this material. This is derived from 172 numerical relativity simulations covering 21 equations of state and mass and compactness ranges of $1<M_i/M_{\odot}<2$ and $0.1 < C_i < 0.23$ where $M_i$ is the mass of the $i$th NS, $R_i$ is its radius and $C_i \equiv G M_i/R_i c^2$ is its compactness \citep{Hotokezaka13,Bauswein13,Dietrich15,Lehner16,Sekiguchi16,Dietrich17a}. The compactness typically depends on the equation of state but in the generative model it is fixed to the value for a NS with $M=1.4$\msol~and radius 10.7 km. While the mass range used in the \texttt{bns\_generative} model does not include some of our higher mass simulated mergers, the results should still be accurate enough for our purposes. \citeauthor{Dietrich17b} found the total dynamical mass to follow
\begin{equation}
\label{eq:m_dyn}
\begin{split}
    \frac{M_{\rm dyn}}{10^{-3}M_{\odot}} = & \left[a\left(\frac{M_2}{M_1}\right)^{1/3}\left(\frac{1-2C_1}{C_1}\right)+b\left(\frac{M_2}{M_1}\right)^{n} \right.\\ & +\left. c\left(1-\frac{M_1}{M_1^*}\right)\right]M_1^* 
    + \left[a\left(\frac{M_1}{M_2}\right)^{1/3}\left(\frac{1-2C_2}{C_2}\right) \right. \\ & \left. +b\left(\frac{M_1}{M_2}\right)^{n} c\left(1-\frac{M_2}{M_2^*}\right)\right]M_2^* + d
\end{split}
\end{equation}
where $M_i^* = M_i + 0.08 M_i^2$ is the baryonic mass of the $i$th NS and $a$, $b$, $c$, $d$ and $n$ are fitted parameters. The velocity of the ejecta is given by
\begin{equation}
\label{v_dyn}
    v = \left[a_v\left(\frac{M_1}{M_2}\right)(1+c_v C_1)\right] + \left[a_v\left(\frac{M_2}{M_1}\right)(1+c_v C_2)\right] + b_v
\end{equation}
where $a_v$, $b_v$ and $c_v$ are fitted parameters with different values in the polar and orbital plane directions.

The dynamical ejecta varies in electron fraction according to its geometry. The equatorial ejecta are dominated by tidal ejecta and have a low electron fraction $Y_e < 0.25$. The material is therefore more capable of producing heavy $r$-process nuclei with high opacity \citep{Barnes13} and hence is termed the red component. In the polar direction, shocks and neutrino heating mean the material has a significantly higher electron fraction $Y_e>0.25$ and can therefore only produce lower mass \textit{r}-process nuclei. The resultant lower opacity leads to bluer emission from this component. In \citetalias{Nicholl21}'s model, the UV/optical opacities of these components are fixed to $\kappa_{\rm red} = 10$ cm$^2$ g$^{-1}$ and $\kappa_{\rm blue} = 0.5$ cm$^2$ g$^{-1}$ following \citet{Radice18}. The ratio between the masses of these red and blue components has been found to depend strongly on the mass ratio between the progenitor NSs \citep{Radice18} and \citetalias{Nicholl21} use the simulations of \citet{Sekiguchi16} to derive the relative masses of each component in each simulated kilonova (e.g. in their Figure 2).

In addition to dynamical ejecta, kilonovae can also be driven by material ejected after the merger driven by winds from the merger remnant and its accretion disk. The merger remnant can take the form of a stable NS, a supermassive NS which eventually collapses to a black hole, or could immediately collapse. NS remnants are predicted to produce larger disk masses with a higher electron fraction and therefore bluer emission \citep{Metzger14b, Lippuner17}. The mass of the disk is found to follow
\begin{equation}
\label{m_disk}
    \log_{10}\left(M_{\rm disk}\right) = \max\left[-1, a\left(1 + b \tanh\left(\frac{c-M_{\rm tot}/M_{\rm th}}{d}\right)\right)\right]
\end{equation}
where $a$, $b$, $c$ and $d$ are fitted parameters, $M_{\rm tot} = M_1 + M_2$. The threshold mass for a prompt collapse to a black hole is then given by
\begin{equation}
\label{eq:m_thresh}
    M_{\rm th} = \left(2.38 - 3.606\frac{M_{\rm TOV}}{R_{1.4}}\right)M_{\rm TOV}
\end{equation}
where $M_{\rm TOV}$ is the Tolman-Oppenheimer-Volkoff mass (the upper mass limit for a non-rotating NS) and $R_{1.4}$ is the radius of a 1.4 \msol~NS. Based on \citetalias{Nicholl21}, we use $M_{\rm TOV} = 2.17$ \msol~and $R_{1.4} = 10.7$ km which yields $M_{\rm th} = 3.58$ \msol.

A fraction of the disk mass, given by $\epsilon_{\rm disk}$ and varying from 0.16 to 0.23, is ejected by viscously driven winds. For our simulations, we therefore draw it from a suitable normal distribution (see Table \ref{tab:kn_params}). \citetalias{Nicholl21} derive the final mass of this component from $M_{\rm tot}$, $M_{\rm TOV}$ and $M_{\rm th}$ following \citet{Metzger19}. The material ejected from the disk is defined as a `purple' component (marked as $M_{\rm purple}$ in Figure \ref{fig:kn_sketch}) and has an intermediate Uv/optical opacity of $\kappa_{\rm purple}$, in turn derived from the simulations of \citet{Lippuner17} and the $Y_e - \kappa$ relations of \citet{Tanaka20} and found to range from 1.5 to 5.5 cm$^2$ g$^{-1}$ depending on the remnant's lifetime.

Winds, specifically those magnetically or neutrino driven, can also induce an additional enhancement in the blue ejecta \citep[e.g.][]{Metzger18,Radice18,Ciolfi20} if the remnant does not promptly collapse, i.e. $M_{\rm tot} < M_{\rm th}$. This is implemented in the model using the parameter $\alpha$, defined as the ratio between the dynamical blue ejecta and total blue ejecta, i.e.$M_{\rm blue, dyn} / M_{\rm blue, tot}$. Setting $\alpha < 1$ therefore provides the enhancement. Of the 922 simulations, 357 (37.8\%) have a total merger mass less than this threshold and we draw a value for $\alpha$ from a uniform distribution between 0.1 and 1.0.

The geometry of all these different components leads to a strong angular dependence. \citetalias{Nicholl21} build on the prescription of \citet{Villar17} using that of \citet{Darbha20}. The kilonova is modelled as a sphere with conical polar caps with half-opening angles of $\theta_{\rm open} = 45\degr$. For a given $\theta_{\rm obs}$, the luminosity of each component is simply scaled using the projected area of the caps for the blue and purple ejecta or remaining sphere for the red ejecta.

The luminosities are derived using the \textsc{MOSFiT} module of \citet{Cowperthwaite17} and \citet{Villar17}. The ejecta expand homologously with a grey opacity and a velocity $v_{\rm ej}$. The temperature of the photosphere is therefore initially calculated from the Stefan-Boltzmann law until the ejecta expand sufficiently to cool and begin to recombine with recombination temperature $T_{\rm rec} = 2500$ K. The luminosity of each component is scaled proportionally to its mass.

A final possible emission component derives from the short GRB itself. As the jet expands, it can shock-heat the surrounding ejecta and form a cocoon. Following \citet{Piro18} and \citet{Nakar17}, \citetalias{Nicholl21} assume the shock deposits constant energy per decade of velocity in the ejecta and therefore the luminosity is proportional to the mass of the shocked ejecta. This is a fraction $\theta_{\rm cocoon}^2/2$ of the total ejecta and for our simulations we draw $\theta_{\rm cocoon}$ from a uniform distribution between 0\degr and 45\degr.

Between the GW simulations and drawing from our selected distributions, we therefore explore a diverse sample of possible kilonovae and effectively evaluate strategies for follow-up. The parameters we use for the simulated light curves are summarised in Table \ref{tab:kn_params}.

\begin{table}
\centering
\caption{The parameter space drawn from to generate the kilonova light curve where $\mathcal{N}(\mu,\sigma^2)$ refers to a normal distribution and $\mathcal{U}(\min,\max)$ refers to a uniform distribution.}
\begin{tabular}{cc}
    \hline
    Parameter & Value\\
    \hline
    Redshift & From GW simulation \\
    $D_L$ & From GW simulation \\
    \thetaobs & From GW simulation \\
    $M_c$ & From GW simulation \\
    $q$ & From GW simulation \\
    $\epsilon_{\rm disk}$ & $\mathcal{N}(0.20,0.03^2)$ \\
    $\alpha$ & $\mathcal{U}(0.1,1.0)$ \\
    $\theta_{\rm cocoon}$ & $\mathcal{U}(0\degr,45\degr)$ \\
    \hline
\end{tabular}
\label{tab:kn_params}
\end{table}

\begin{figure}
    \centering
    \includegraphics[width=\columnwidth]{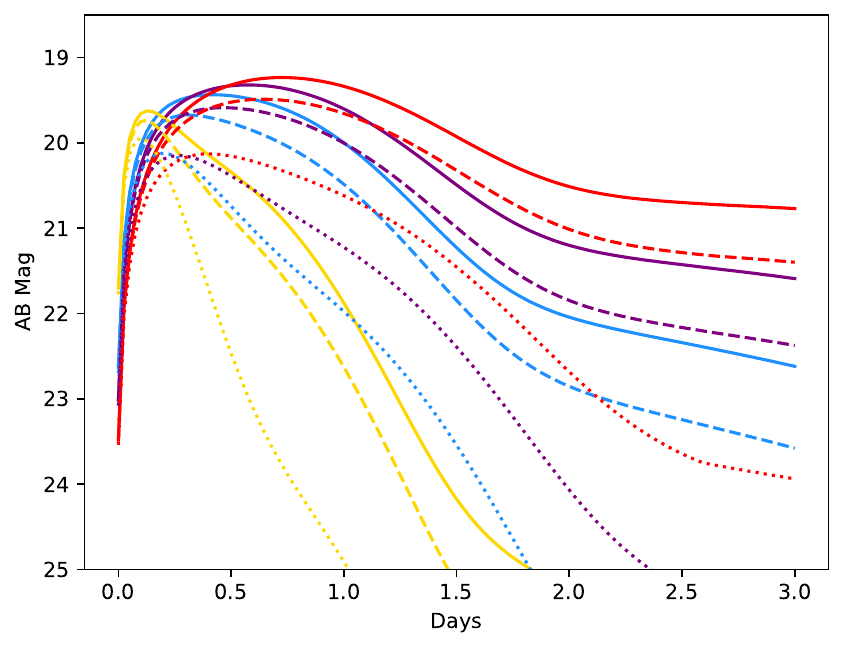}
    \caption{The \textit{uvm2} (yellow), \textit{u} band (blue), \textit{b} band (purple), \textit{v} band (red) light curves of an example kilonova at 100 Mpc, driven by the merger of two 1.4 \msol~NSs. It is viewed at three observation angles, 0\degr (solid), 45\degr (dashed) and 90\degr (dotted).}
    \label{fig:example_kn}
\end{figure}

\subsection{Afterglow component}
\label{sec:afterglow_model}

There is also the possibility of significant afterglow contribution to the light curves at UV wavelengths. This, however, has a much stronger dependency on \thetaobs~than the kilonova component.

We evaluate the afterglow contribution using the semi-analytic code \textsc{afterglowpy v.0.7.3} \citep{Ryan20} which uses the single shell approximation of \citet{vanEerten10} and \citet{vanEerten18}. In this model, the ejecta, contact discontinuity and forward shock are integrated over and treated as a single radially uniform fluid element. To interpolate between the ultra-relativistic and non-relativistic regimes, a trans-relativistic equation of state is employed \citep{vanEerten13,Nava13}. The final model is calibrated to the \textsc{BoxFit} code \citep{vanEerten12}. This uses full high resolution 2D numerical hydrodynamics simulations of jets to derive afterglows.

Traditionally, GRB jets are assumed to be `top hat' jets, with roughly constant energy across the jet and a precipitous decline at its edge. However, it is likely that jets have significantly more structure and \textsc{afterglowpy} includes several different possibilities. 
For instance, both Gaussian and power law structures are modelled. In our case, we assume the jets to have a Gaussian structure, as was likely the case for GW 170817 \citep[e.g.][]{Lamb19a,Salafia19,Troja19,Ryan20}, and therefore the energy of the jet varies with angle $\theta$ as 
\begin{equation}
    E(\theta) = E_0 \exp \left( - \frac{\theta^2}{2\theta_c^2} \right)
    \label{eq:gaussian}
\end{equation}
where we have taken $\theta_c$ as the half opening angle of the jet\footnote{We note that this is a slightly different assumption than that made in \citet{Ryan20} and may result in the afterglow contribution being slightly overestimated. However, the effect of this is minimal as the vast majority of sources are still dominated by kilonova emission in the \textit{u} band.} i.e. $\theta_c = \theta_j/2$ and the truncation angle as $\theta_w = 4 \theta_c$ to a maximum of $45 \degr$. To model a structured jet, \textsc{afterglowpy} breaks it down into many top hat components. The properties of the blast wave for each of these components, primarily the radial position of the shock, dimensionless four-velocity and time-dependent jet opening angle, are described by a set of ordinary differential equations. These are numerically integrated and solved and the top hat components are summed over to derive the resultant afterglow of the GRB.

For each simulation, some input parameters to the afterglow model can be taken from the GW simulation, primarily the observing conditions. The remaining parameters of the short GRB powering the afterglow are drawn from the distributions observed in the \textit{Swift} SGRB population \citep{Fong15}. This parameter space is summarised in Table \ref{tab:grb_params}. Again, drawing from a range of parameters allows a properly diverse sample of afterglow contributions to be evaluated and our observing strategy to be fully optimised.

\begin{table}
\centering
\caption{The parameter space drawn from to generate afterglow contributions to the light curve where $\mathcal{N}(\mu,\sigma^2)$ refers to a normal distribution. Note that $\theta_j$ is the jet opening angle.}
\begin{threeparttable}
\begin{tabular}{cc}
    \hline
    Parameter & Value\\
    \hline
    Redshift & From GW simulation \\
    $D_L$ & From GW simulation \\
    \thetaobs & From GW simulation \\
    $\log(\meiso)$ & $\mathcal{N}(51.3,0.7^2)$ erg\\
    $\theta_j$ & $\mathcal{N}(16,10^2) \degr$ \\
    $p$ & $\mathcal{N}(2.37,0.25^2)$ \\
    $\log(n_0)^{\dagger}$ & $\begin{cases} P=0.57, \mathcal{N}(-3.81,0.43^2) \\ P=0.43, \mathcal{N}(-0.02,0.92^2)\end{cases}$ cm$^{-3}$ \\
    $\epsilon_e$ & 0.1 \\
    $\epsilon_B$ & 0.01 \\
    \hline
\end{tabular}
\begin{tablenotes}
\item $^{\dagger}$ $\log(n_0)$ has a bimodal distribution, which we model as two normal distributions with relative probabilities of being drawn from of 0.57 and 0.43. 
\end{tablenotes}
\end{threeparttable}
\label{tab:grb_params}
\end{table}

\subsection{\textit{Swift}'s instrumental response}
\label{sec:instrumental_response}

The next step is to combine the individual components from our kilonova and afterglow modelling in Section \ref{sec:lc_modelling} and calculate the responses from \textit{Swift}'s XRT and UVOT. For the XRT, we assume the contribution from any kilonova would be negligible and therefore only the afterglow emission would be relevant. The range of spectra we model are similar enough that a simple conversion from the 1 keV flux density calculated by \textsc{afterglowpy} 
yields sufficiently representative count rates (CR) in the full XRT band. This conversion, which assumes $N_{\rm H} = 3\times10^{20}$ cm$^{-2}$ and $\Gamma=2$, gives for the flux density in Jy,
\begin{equation}
    {\rm CR}_{\rm 0.3-10\,keV} = S_{\rm 1\,keV} \times 2.07\times10^5\,{\rm ct\, s}^{-1}.
\end{equation}
For the UVOT, we calculate the kilonova and afterglow components for the \textit{u}, \textit{b}, \textit{v}, \textit{uvw1}, \textit{uvm2} and \textit{uvw2} bands. We convert the AB magnitudes computed by \textsc{afterglowpy} and \textsc{MOSFiT} to count rates using each filter's AB zero point \citep{Breeveld11} and apply an additional calibration factor $\lesssim 1$ to account for the decrease in sensitivity of UVOT over \textit{Swift}'s life\footnote{\url{https://heasarc.gsfc.nasa.gov/docs/heasarc/caldb/swift/docs/uvot/uvotcaldb_throughput_07.pdf}}. To derive the total number of counts for each simulation, we multiply the total count rate for both the kilonova and the afterglow components by the exposure time.

It is relatively straightforward to determine if a source is likely to be detected by the XRT, with a threshold of six counts.  However, it is more complicated for UVOT as the background is significantly greater and can vary by more than a factor of two, depending on Galactic latitude, ecliptic latitude and Earth or Sun limb angle. To fully model the UVOT's response, we use example images of exposure times approximately equal ($\pm 15$ s) to those used in our strategy. These exposures are selected from those far outside the Galactic plane to avoid diffuse Galactic background or Zodiacal contamination, and are also chosen so that we had a `median' background and `high' background to test. Artificial sources are injected into the UVOT images with their total number of counts taken from the results of our simulations. This is repeated 25 times for each simulation then the \textsc{uvotdetect} utility was used to extract the sources. When running \textsc{uvotdetect}, \texttt{zerobkg} is set to -1, which calculates the local background for all points in the image, and the detection threshold is set to 2-$\sigma$ as these are the typical parameters used by the UVOT team. The fraction of the 25 injected artificial sources recovered is then taken to be the detection probability.

\section{Optimising \textit{Swift}'s observing strategy}
\label{sec:obs_strategy}

There are a number of factors we can control in terms of \textit{Swift}'s observing strategy. In terms of the observations themselves, it is important to select the filter and exposure time correctly. It is also crucial to use \textit{Swift} efficiently for all science, not just looking for EM counterparts to LVK detections, in particular using \textit{Swift}'s unique UV capability. We must therefore ensure we focus on those triggers where \textit{Swift} can be effective. There are three pieces of information that LVK release that can help us make this decision - the object type probabilities, the distance to the event and the area of 90\% error region. In this section, we explore each factor that affects our strategy in turn.

\subsection{Filter selection}

\begin{figure*}[ht!]
    \centering
    \includegraphics[width=\columnwidth]{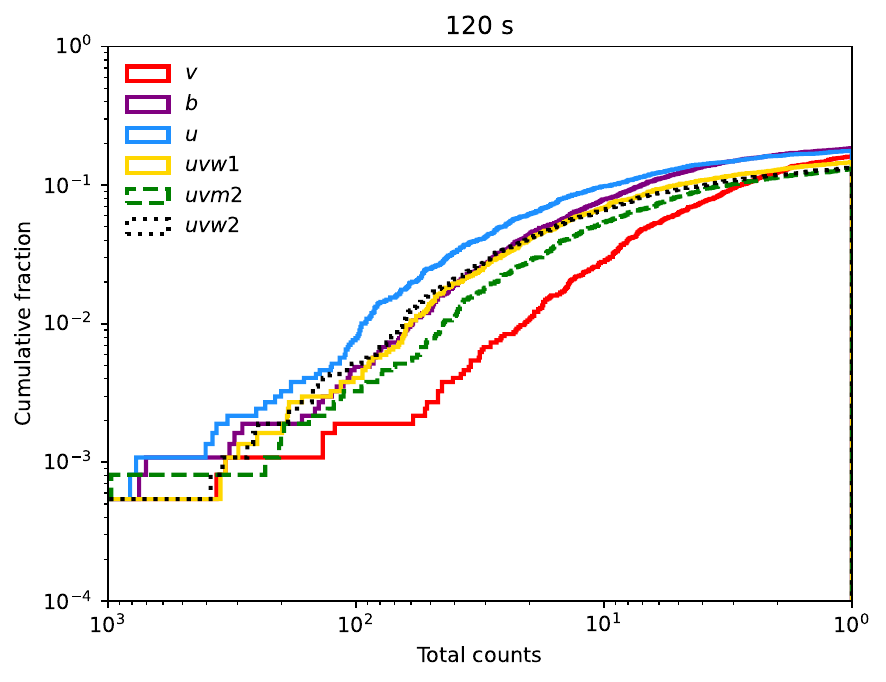}
    \includegraphics[width=\columnwidth]{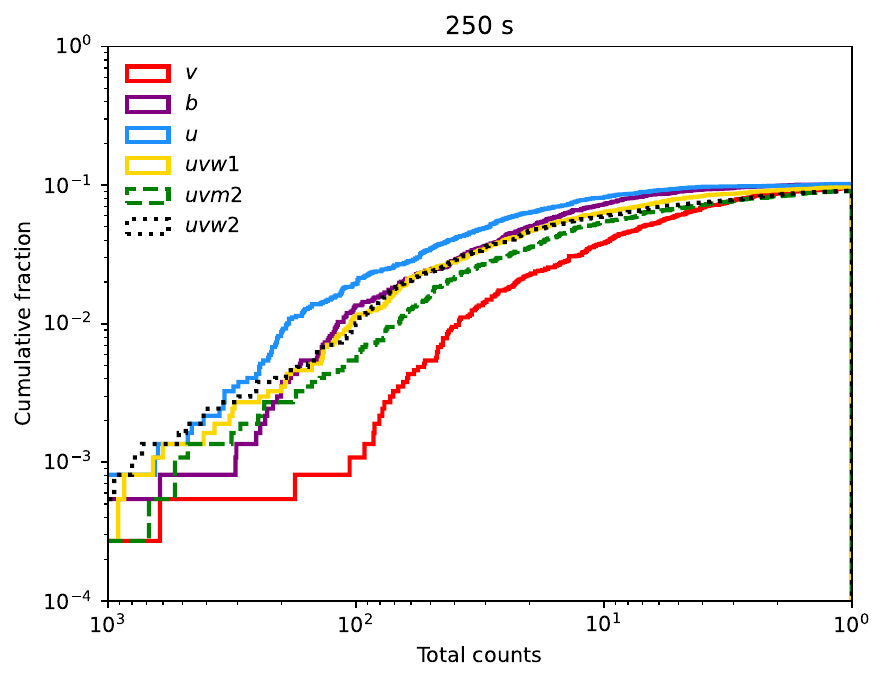}
    \includegraphics[width=\columnwidth]{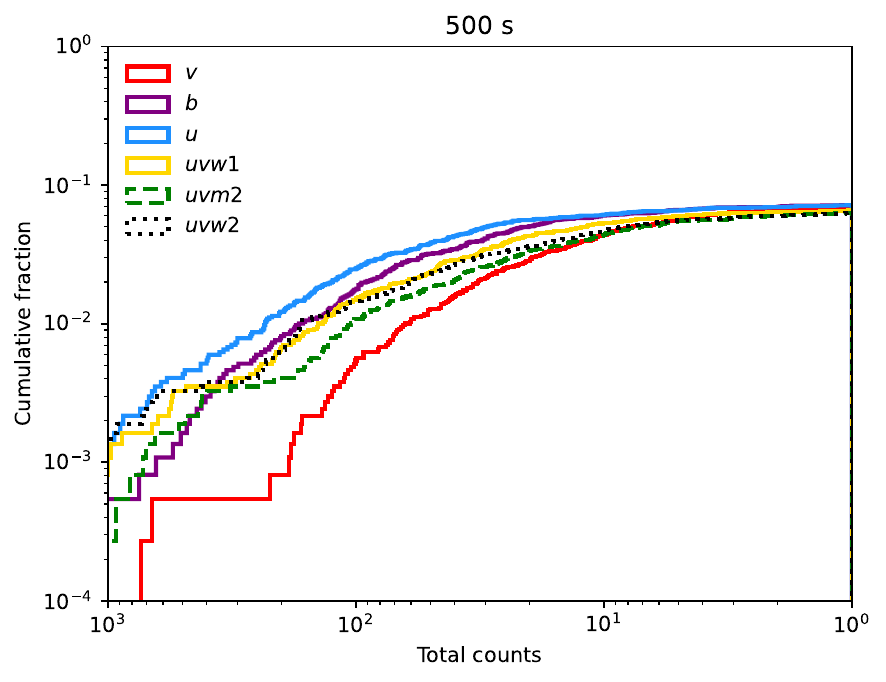}
    \includegraphics[width=\columnwidth]{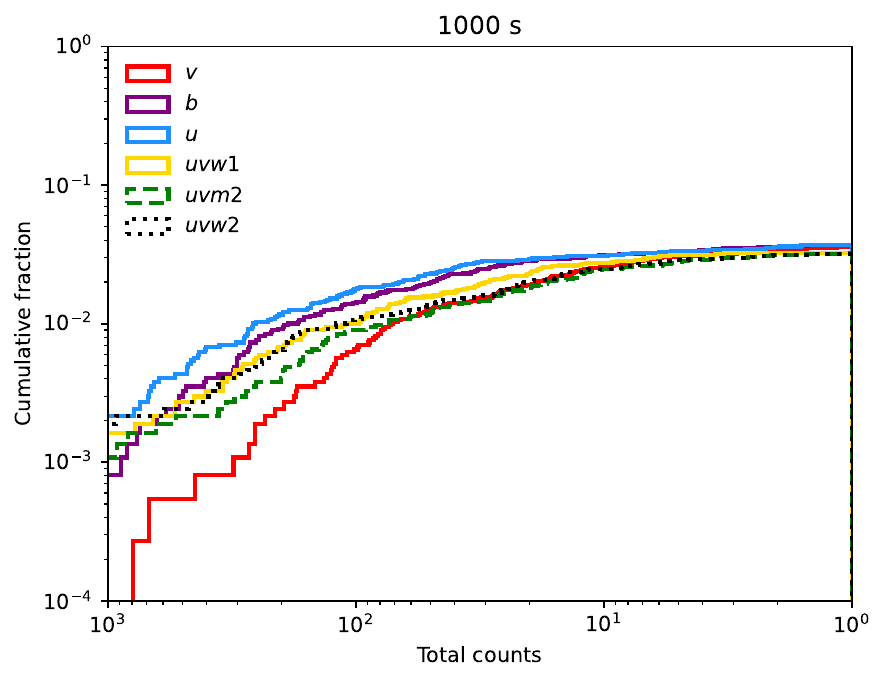}
    \caption{Results from our simulations, the cumulative fraction of simulated triggers against total counts. The title of each panel refers to the exposure time for each observation i.e. $t_{\rm exp}$. Note the scale on the x axis goes from large to small.}
    \label{fig:preliminary_results}
\end{figure*}

Kilonovae are primarily thermally driven and peak at relatively low temperatures. The vast majority of events will be seen at high observation angles and therefore be very `red', with fast colour evolution away from blue, as seen in Figure \ref{fig:example_kn}, although there will also be a significant fraction of events that will be viewed relatively `on-axis' and therefore appear `blue' for an appreciable period of time. The UVOT is optimised for bluer sources, likely to be the minority of triggers particularly at later times, and not wide areas. This means that ensuring \textit{Swift} reaches the triggers on a rapid timescale is more important than, for instance, a wide field optical instrument.

From our preliminary results, we find the \textit{u}, \textit{b}, \textit{uvw1}, \textit{uvm2} and \textit{uvw2} counts are broadly comparable at early times for various strategies, as shown in Figure \ref{fig:preliminary_results}, and consistent with the example kilonova light curve in Figure \ref{fig:example_kn}. In the \textit{v} band, the counts are significantly lower. We therefore discount this band for our follow-up. We found this trend was consistent across all our simulations, regardless of the properties of the individual GW event. 

We also needed to consider the background and therefore likely signal to noise in each filter. The background of the \textit{b} and \textit{v} bands is significantly greater than that of the \textit{u}, \textit{uvw1}, \textit{uvm2} and \textit{uvw2} bands \citep{Breeveld10}. As the \textit{u} band has a small apparent advantage in Figure \ref{fig:preliminary_results} and the \textit{Swift} Gravitational Wave Galaxy Survey \citep{Tohuvavohu19} will provide image subtraction templates to aid in identification of new sources, we therefore concentrate on it for our follow-up strategy. Finally, the NUV  wavelengths covered by the \textit{u} band \citep[$\sim3500$ \AA, ][]{Poole08} cannot be covered by ground-based telescopes which are, in turn, better optimised than \textit{Swift} to search in the visible filters. For the \textit{u} band, the median and high backgrounds described in Section \ref{sec:instrumental_response} are 0.01 ct s$^{-1}$ pixel$^{-1}$ and 0.015 ct s$^{-1}$ pixel$^{-1}$. We only use the procedure detailed there for sources where the expected number of counts is at least 10, however, as fewer counts are essentially undetectable.

\subsection{Exposure time}
\label{sec:exposure_times}

The biggest factor in optimising \textit{Swift}'s strategy is the time taken observing each field - the exposure time. It impacts two things - the time it takes to get to the correct field (see Equation \ref{eq:t_reach}) and how likely we are to detect the source when it is reached.

To simplify the process of responding to triggers and the required controlling, we concentrate on strategies with a fixed exposure time per field. These exposure times are 120 s, 250 s, 500 s and 1000 s with 120 s being approximately the minimum feasible time for a \textit{Swift} observation. We did briefly examine strategies where the exposure time varied based on the time since the trigger but found no significant differences compared with the simpler strategy.

\subsubsection{$t_{\rm reach}$}

\begin{figure}
    \centering
    \includegraphics[width=\columnwidth]{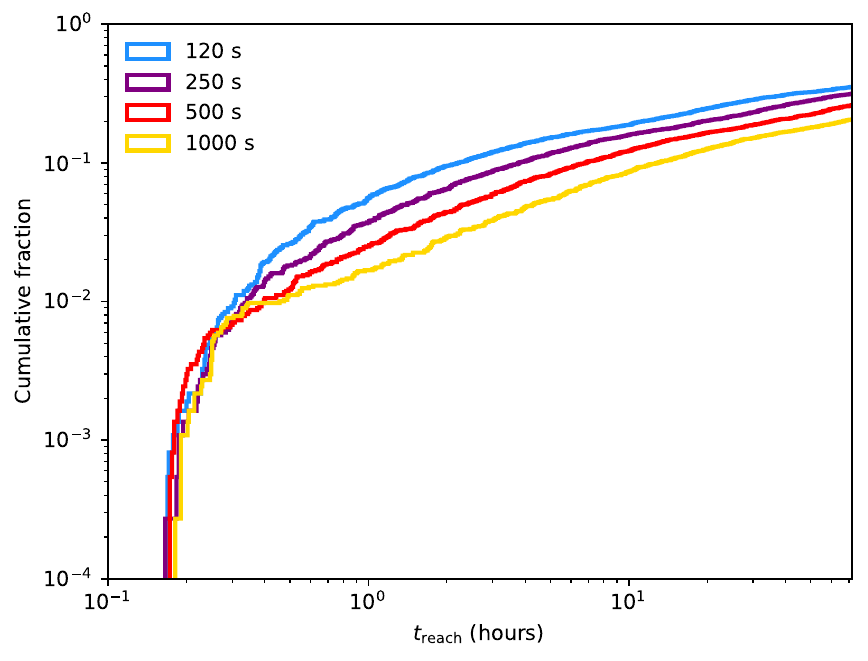}
    \caption{The cumulative fraction of the 3688 simulated triggers where \textit{Swift} reaches the correct field against $t_{\rm reach}$ for the first 72 hours following the trigger for each exposure time.}
    \label{fig:t_reach}
\end{figure}

From the list of fields for each simulation, we calculated the $t_{\rm reach}$ from Equation \ref{eq:t_reach} for each possible $t_{\rm exp}$. The cumulative distribution for the simulations is shown in Figure \ref{fig:t_reach}, where we set an upper limit of 72 hours as by this time, any EM signals are too faint for \textit{Swift} to detect in the exposure times being probed here.

\begin{figure}
    \centering
    \includegraphics[width=\columnwidth]{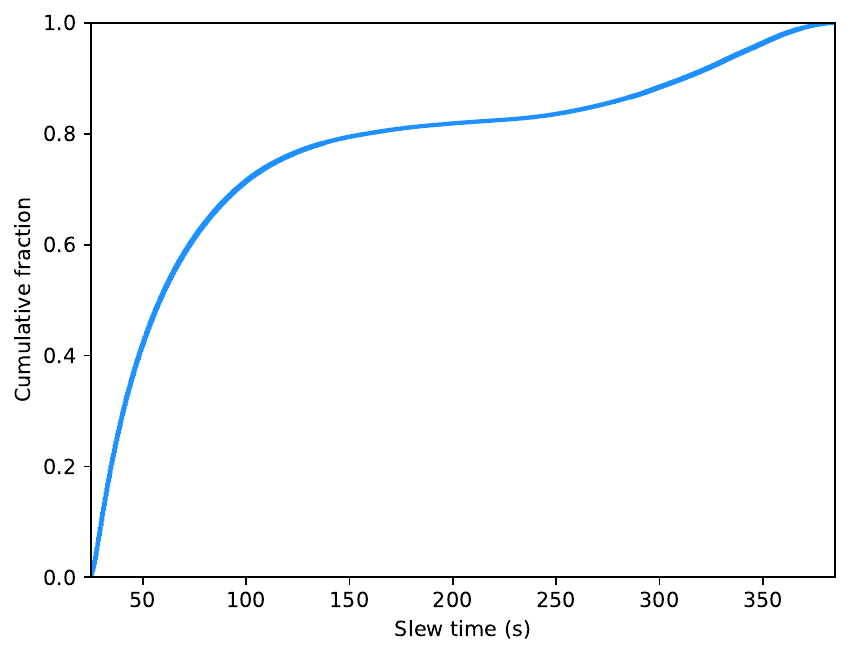}
    \caption{The cumulative fraction of individual slews against their length.}
    \label{fig:slew_times}
\end{figure}

Unsurprisingly, the 120 s exposure strategy reaches the correct fields most rapidly achieving this within 24 hours for 26.5\% of our simulations. However, the equivalent fractions for the 250 s, 500 s and 1000 s are 21.3\%, 17.3\% and 13.6\%, higher than might be expected if the exposure time was the dominant factor in $t_{\rm reach}$. Instead, the total slew time dominates. We plot the distribution of individual slew times in Figure \ref{fig:slew_times}. The majority of slews are reasonably short ($\sim 50$\% are less than 60 s and $\sim 80$\% are less than 150 s) but there is a significant fraction of slews that take hundreds of seconds. This is likely a result of skymaps with disparate regions of high probability. While the field-selection algorithm does de-weight long slews, maximising the probability covered while accounting for observing constraints means they are sometimes necessary. However, because often thousands of fields must be observed before \textit{Swift} reaches the correct location, the number of slews dominates over their individual lengths.

\subsubsection{Counterpart detection}

\begin{figure}
    \centering
    \includegraphics[width=\columnwidth]{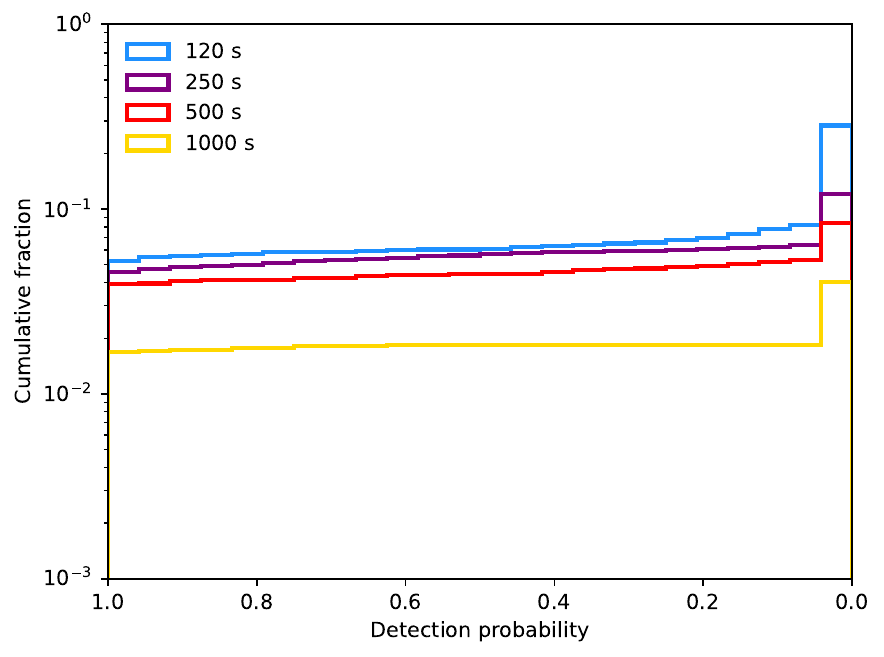}
    \includegraphics[width=\columnwidth]{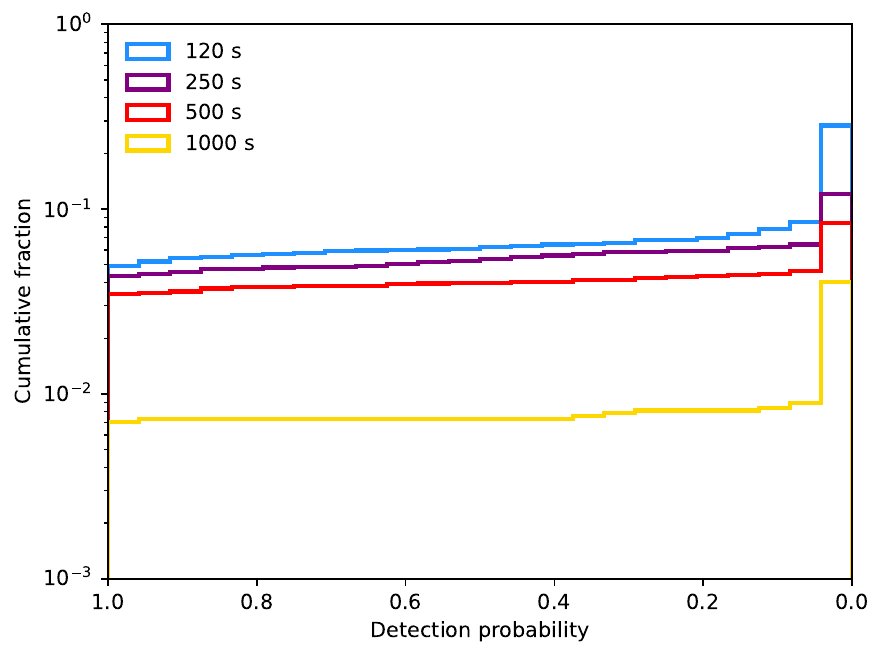}
    \caption{The cumulative fraction of simulated triggers against detection probability in the \textit{u} band for the median background (top) and higher background (bottom). Note the x axis goes from large to small and that the y axis is in log scale.}
    \label{fig:detprob_all}
\end{figure}

We must also consider how likely each instrument is able to detect a given source. Their\footnote{We note that this isn't necessarily the case for the XRT which tends to be photon dominated at these relatively short exposures and hence the sensitivity varies roughly as $t$.} sensitivities vary as approximately $t_{\rm exp}^{1/2}$, hence, for slowly evolving light curve components, the exposure time will have a greater impact than the time to reach the field containing the source. In Figure \ref{fig:detprob_all}, we show the cumulative distribution of the 3688 simulated triggers against their detection probability. For both backgrounds, we find that 120, 250 and 500 s are roughly comparable with 1000 s significantly worse. This is probably because for the 1000 s strategy, the sources are inherently fainter having been reached at a later time, and therefore the signal to noise when observed is significantly lower. From our results, it appears that 120 s is the optimal time for the  majority of triggers. We note that this is subject to small number statistics with only a relatively small sample of suitable simulated events.

Our results also show that for the 120 s strategy, $\sim 5.7$\% of triggers have a detection probability $\geq 0.5$. Because it is impractical for \textit{Swift} to follow-up every trigger due to its other observing duties, it is therefore crucial to focus our follow-up on those sources whose properties maximise our chance of detection\footnote{We note that \textit{Swift} does already have such criteria in place, however, we are refining and adding to these criteria here.}.

\subsection{Trigger selection}

The notices issued by LVK when a trigger is detected provide valuable information on whether or not to follow-up. For instance, we can derive the probability that the event contains a disrupted NS:
\begin{equation}
    P_{\rm DNS} = \texttt{PROB\_REMNANT} \times ( 1 - \texttt{PROB\_TERRES})
\end{equation}
where \texttt{PROB\_REMNANT} and \texttt{PROB\_TERRES} are the probabilities of a merger remnant or that the trigger was terrestrial (noise)\footnote{See \url{https://emfollow.docs.ligo.org/userguide/index.html} for details of these parameters.}. These are taken directly from the notices. If $P_{\rm DNS}$ is low, the chance of an EM counterpart, particularly a bright one, is accordingly small and therefore only the best candidates are likely to be worth following up.

\begin{figure*}
    \centering
    \includegraphics[width=\columnwidth]{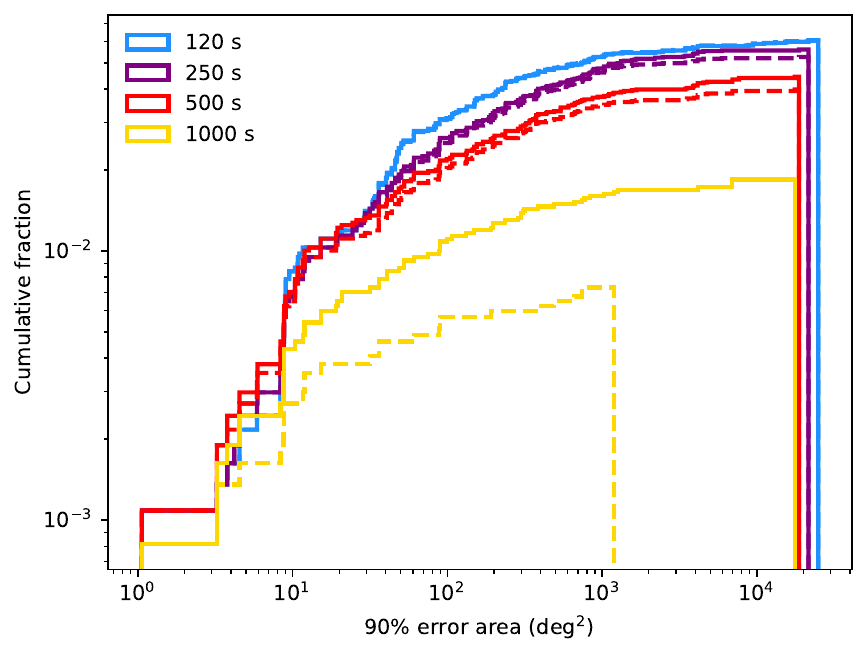}
    \includegraphics[width=\columnwidth]{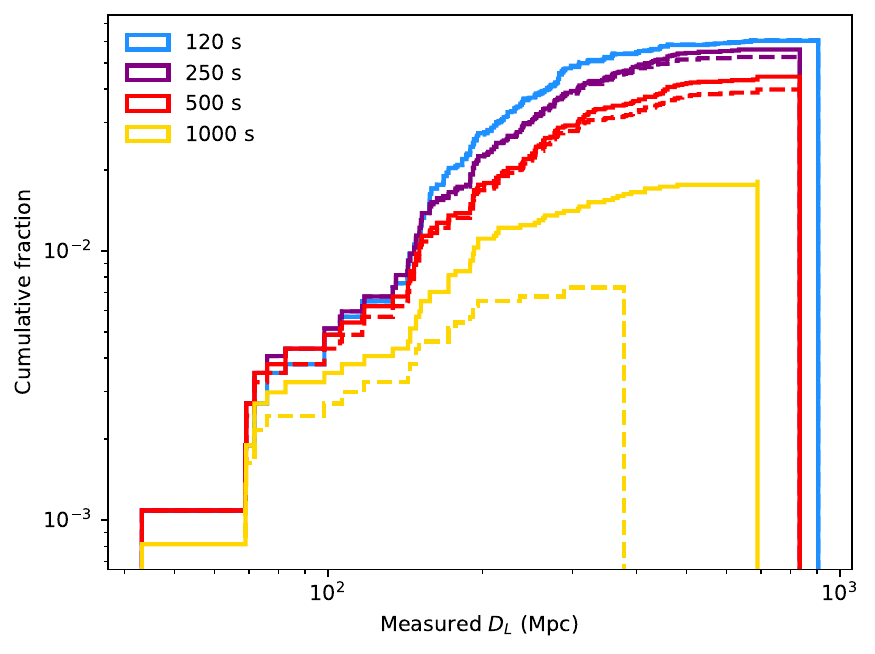}
    \caption{The cumulative fraction of simulated triggers with detection probability $\geq 0.5$ against 90\% error area (left panel) and measured $D_L$ (right panel). The median and higher backgrounds are indicated with solid and dashed lines respectively. Note the axes are in log scale.}
    \label{fig:skymap_summary}
\end{figure*}

In addition, they include a skymap encoding both probability and distance information. The skymaps are processed by our pipeline and the 90\% probability area is extracted. From Figure \ref{fig:skymap_summary}, it is clear that both probability area and distance have significant impacts on how likely \textit{Swift} is to detect a counterpart. Here, we therefore examine in greater detail how the area and distance affect how the EM sources are recovered and which triggers we should follow-up.

\subsubsection{90\% error area}

We binned the simulated skymaps based on the 90\% error area with bins of <150 \degsq, 150 to 300 \degsq, 300 to 500 \degsq, 500 to 1000 \degsq and 1000 to 5000 \degsq. 

\begin{figure*}
    \centering
    \includegraphics[width=\columnwidth]{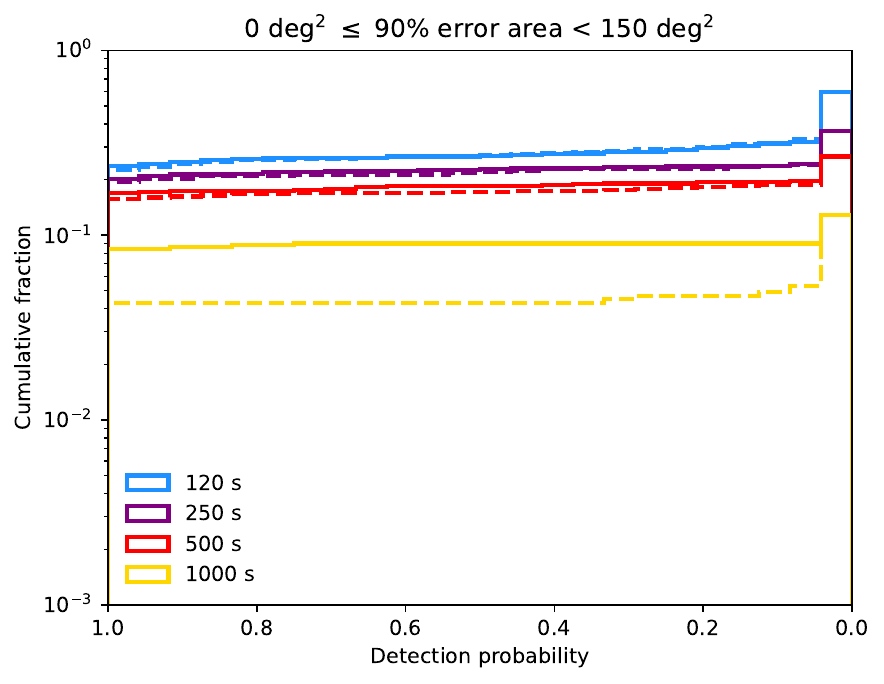}
    \includegraphics[width=\columnwidth]{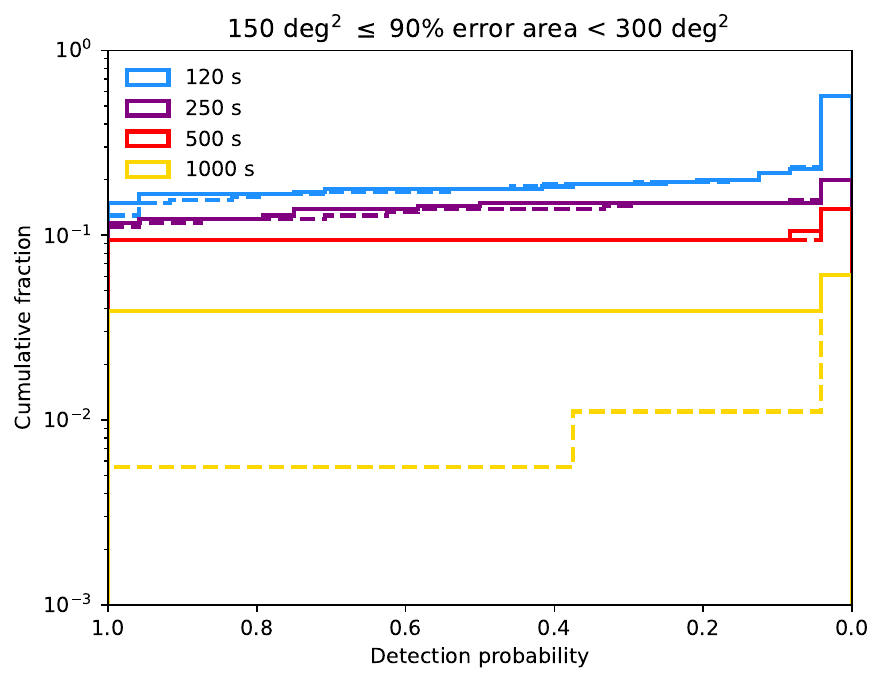}
    \includegraphics[width=\columnwidth]{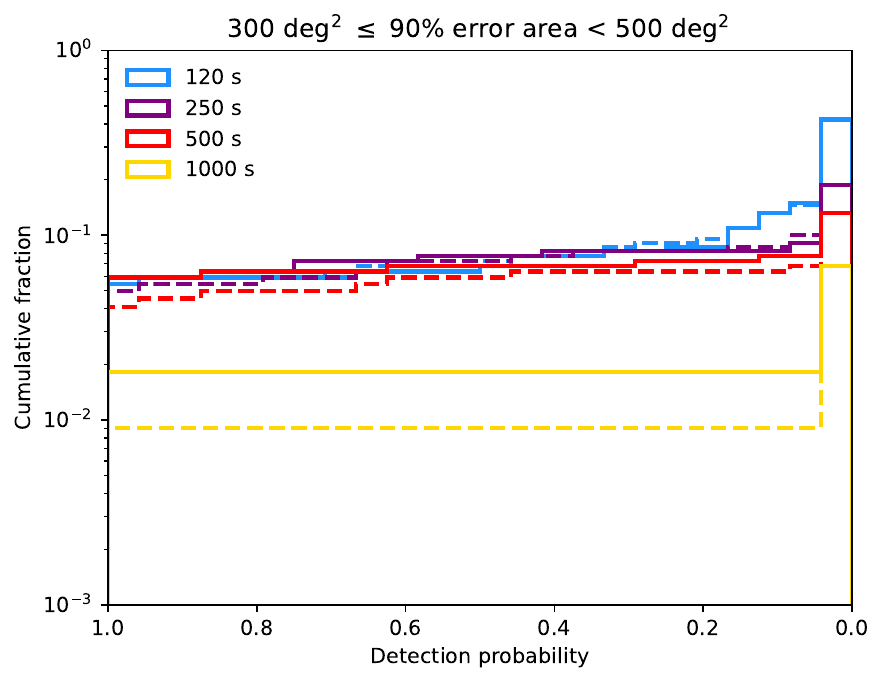}
    \includegraphics[width=\columnwidth]{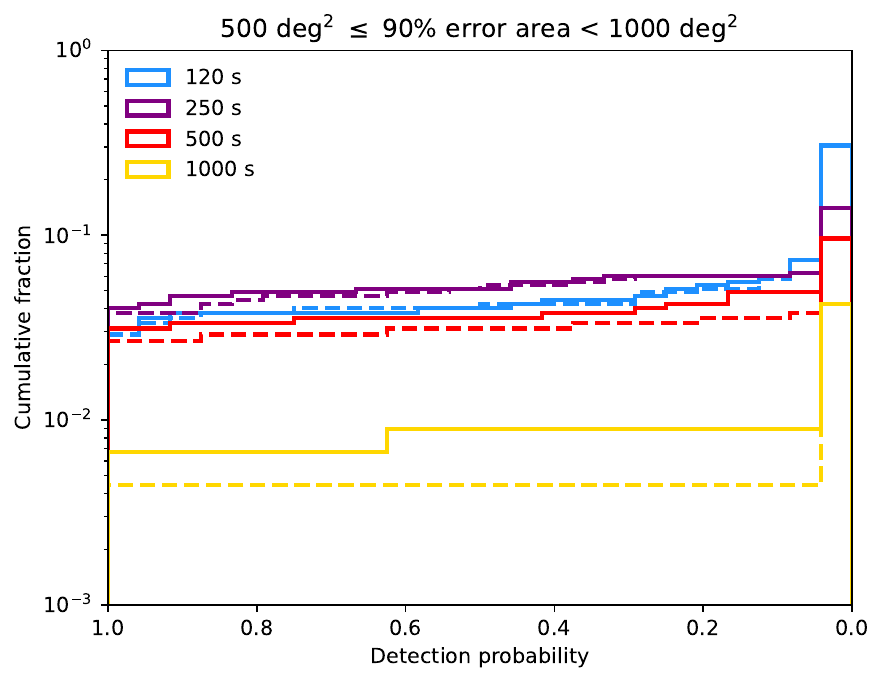}
    \includegraphics[width=\columnwidth]{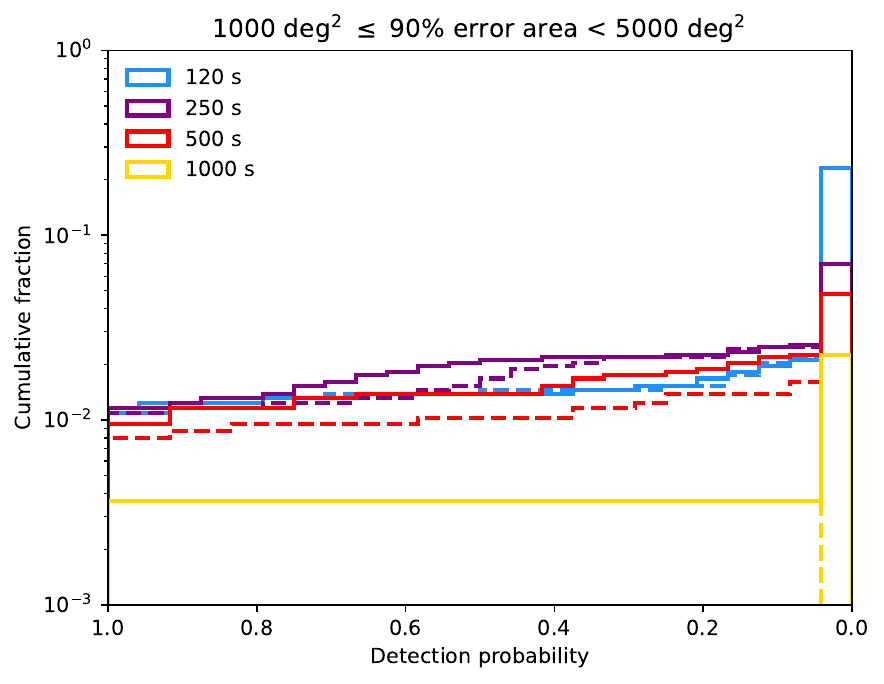}
    \caption{The cumulative fraction of simulated triggers against detection probability in the \textit{u} band for triggers with different 90\% error areas as indicated by the panel titles. The median and higher backgrounds are indicated with solid and dashed lines respectively. Note the x axis goes from large to small and that the y axis is in log scale.}
    \label{fig:area90}
\end{figure*}

In Figure \ref{fig:area90}, we plot the cumulative fraction of simulated triggers against detection probability for each bin. We are again subject to small number statistics, however, for very well localised sources ($\lesssim 150$ \degsq), the fraction of sources that have detection probabilities greater than 0.5 is of order 20 -- 25\% for both 120 and 250 s strategies. For larger areas ($\lesssim 300$ \degsq), there is a decrease but we might expect around 10\% of triggers to have detectable counterparts. The probability of \textit{Swift} detecting a counterpart does continue to decline, however, there is a non-negligible chance of recovering a counterpart\footnote{Although we note that other instruments are much more optimised to tackle a search of this nature.} even with a sky area of order 1000 \degsq. Our results do show that 120 s is an optimal strategy for sources with smaller 90\% error areas, however at $\sim300$ \degsq, the greater sensitivity achieved in a 250 s exposure means this strategy is more suitable.

During O4a, only the two LIGO detectors were online and as at least three detectors are required to achieve a localisation better than $\sim$1000 \degsq, \textit{Swift} did not follow up any triggers. This is a significant issue that affects all narrow-field telescopes including the UVOT. However, the addition of Virgo during O4b has greatly improved the localisation. For instance, the candidate NSBH trigger S240422ed \citep{LVK24a} had a 90\% error area of only 250 \degsq\,and was therefore followed up by \textit{Swift}. We do note, however, that S240422ed has been subsequently downgraded and is most likely terrestrial. Nevertheless, it implies that more suitable triggers will be detected when more GW detectors are available.

\subsubsection{Distance}

\begin{figure*}
    \centering
    \includegraphics[width=\columnwidth]{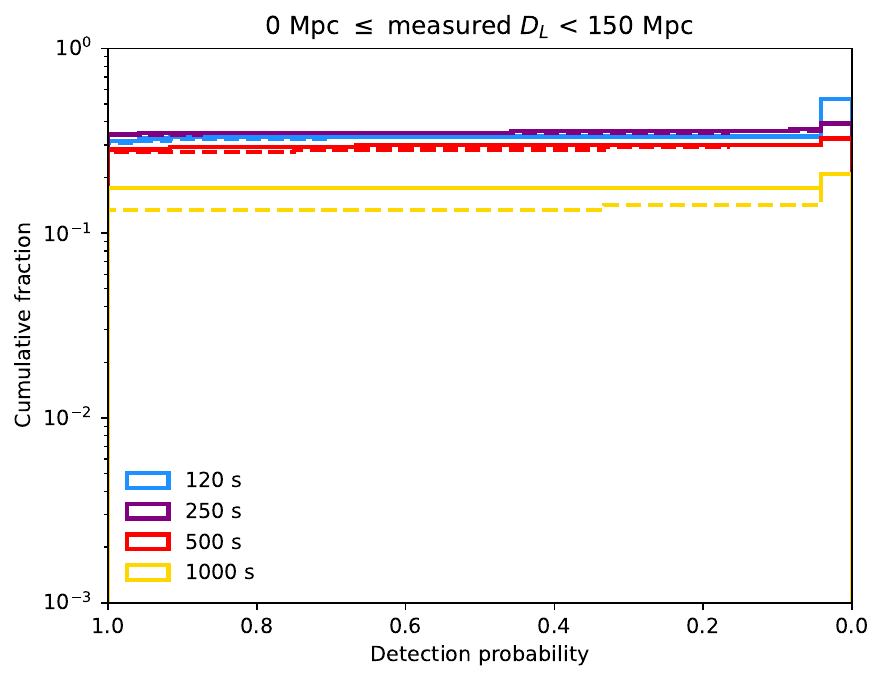}
    \includegraphics[width=\columnwidth]{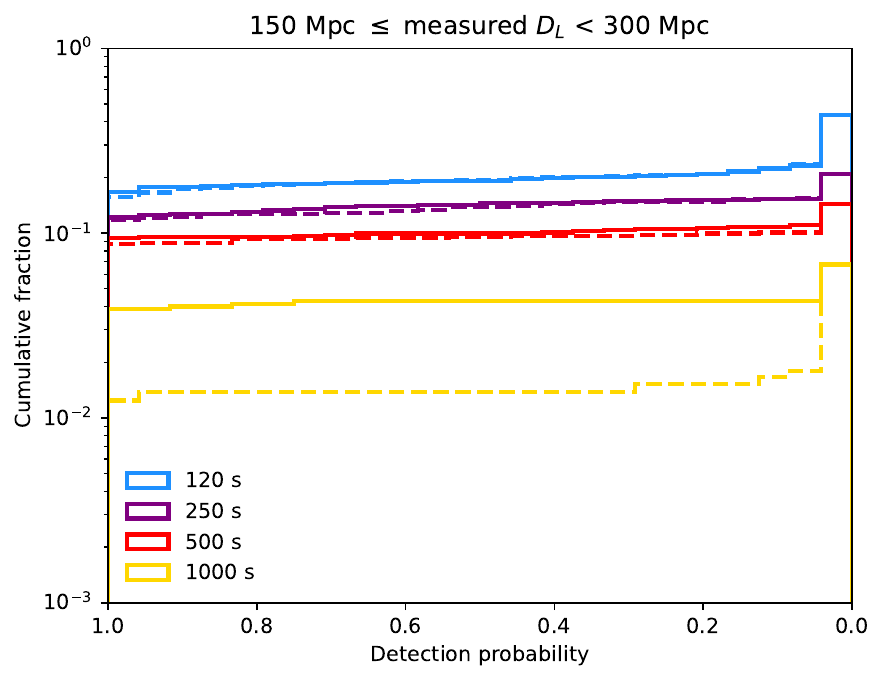}
    \includegraphics[width=\columnwidth]{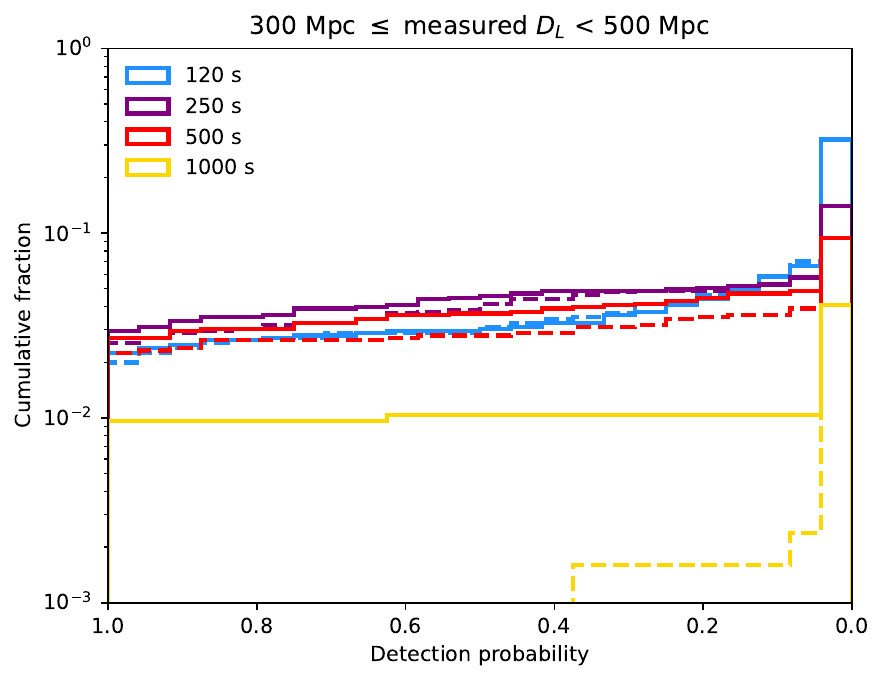}
    \includegraphics[width=\columnwidth]{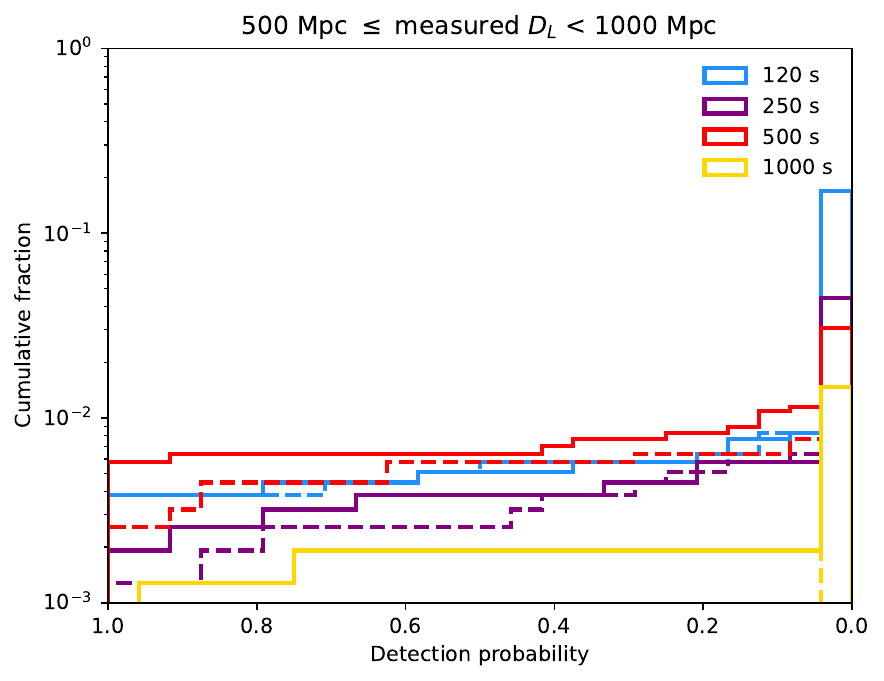}
    \caption{The cumulative fraction of simulated triggers against detection probability in the \textit{u} band for triggers with different measured distances as indicated by the panel titles. The median and higher backgrounds are indicated with solid and dashed lines respectively. Note the x axis goes from large to small and that the y axis is in log scale.}
    \label{fig:measuredD}
\end{figure*}

The distance to an event will also have a significant impact. While the `real' distances to each simulated trigger are known, for actual events this will not be the case; we therefore investigate detection as a function of the calculated distance reported via the \texttt{DISTMEAN} keyword in the skymap.

In Figure \ref{fig:measuredD}, we plot the cumulative fraction of triggers against detection probability for four measured distance bins: <150 Mpc, 150 to 300 Mpc, 300 to 500 Mpc and 500 to 1000 Mpc. At lower distances ($\lesssim 300$ Mpc) our results are consistent with an optimal $t_{\rm exp}$ of 120 s and we might expect 20 to 30\% of triggers resulting in detections. At greater distances up to $\sim$500 Mpc, a 250 s strategy appears more optimal but there is a significant decrease in the probability of a successful detection to a few percent. Beyond this point the probability of a detection is very low and attempting to follow up these sources is inadvisable. We note that it appears that distance has a greater impact on \textit{Swift}'s discovery ability than the sky localisation. There is also likely to be a contributory factor from the completeness of the chosen galaxy catalogue. However, as shown by \citetalias{Evans16b}, 2MPZ is relatively complete over the distances relevant here and while newer catalogues have been released \citep[e.g. GLADE+,][]{Dalya22}, the additional completeness is not overly significant.

\subsubsection{Combined error area and distance}

\begin{figure*}
    \centering
    \includegraphics[width=\textwidth]{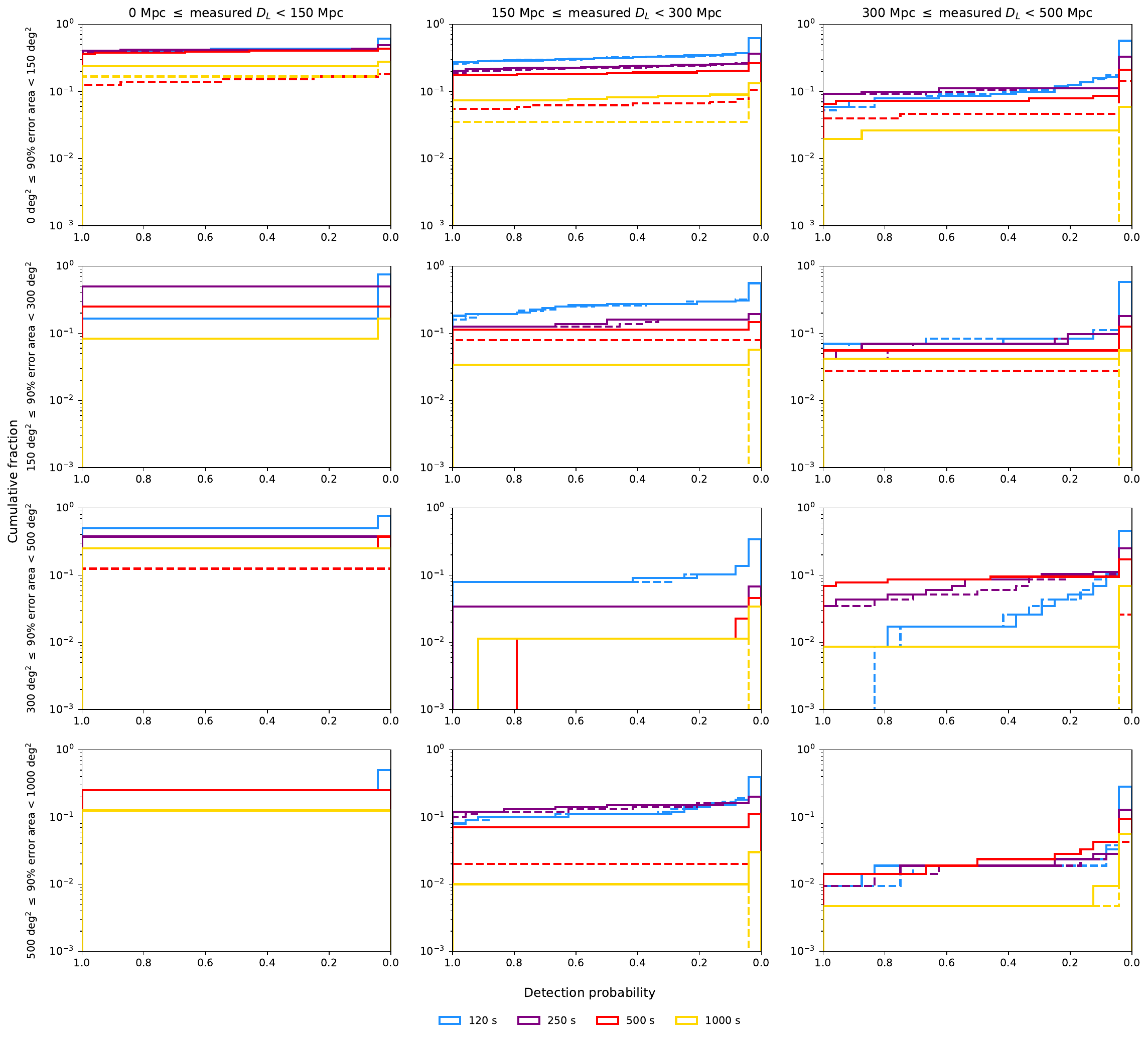}
    \caption{The cumulative fraction of simulated triggers against detection probability in the \textit{u} band for triggers with different 90\% error areas and measured luminosity distances as indicated by the row and column labels respectively. The median and higher backgrounds are indicated with solid and dashed lines respectively. Note the x axis goes from large to small.}
    \label{fig:grid}
\end{figure*}

\begin{table*}
    \centering
    \caption{The fraction of simulated triggers with detection probability $\geq 0.5$ in each measured $D_L$ and 90\% error area bin for 120 s / 250 s / 500 s exposure times and a median background. Note that these bins are highly vulnerable to small number statistics, particular at low measured $D_L$ e.g. the 0 - 150 Mpc / 300 - 500 \degsq bin.}
    \begin{tabular}{c|ccc}
    \diagbox{90\% error area}{Measured $D_L$} & 0 - 150 Mpc & 150 - 300 Mpc & 300 - 500 Mpc \\ \hline
    0 - 150 \degsq & 0.43 / 0.42 / 0.39 & 0.31 / 0.23 / 0.18 & 0.09 / 0.11 / 0.07 \\
    150 - 300  \degsq & 0.17 / 0.50 / 0.25 & 0.26 / 0.14 / 0.11 & 0.07 / 0.07 / 0.06 \\
    300 - 500  \degsq & 0.50 / 0.38 / 0.25 & 0.08 / 0.03 / 0.01 & 0.02 / 0.09 / 0.09 \\
    500 - 1000 \degsq & 0.25 / 0.25 / 0.25 & 0.11 / 0.14 / 0.07 & 0.02 / 0.02 / 0.02 \\
    \end{tabular}
    \label{tab:grid}
\end{table*}

While independently, error area and distance can provide some information on whether to follow-up an event, it is necessary to examine both to fully assess the trigger. In particular, the error area and distance are related - a more nearby trigger is more likely to result in better localisation.

In Figure \ref{fig:grid}, we plot a grid of distance against the sky localisation region size. We also compare the fraction of triggers with detection probability $\geq 0.5$ for 120 s and 250 s exposure times in Table \ref{tab:grid}. As our results above showed, the recovery fraction for triggers in the largest area and distance bins is negligible and we therefore do not examine it here. While each bin is especially susceptible to small number statistics here, we can draw some conclusions. In particular, there appears to be a $\geq$25\% chance of detecting a trigger's EM counterpart if it is within 150 Mpc irrespective of the error area. For luminosity distances between 150 and 300 Mpc, the chances are somewhat lower and decrease further for areas greater than 300 \degsq. At distances greater than 300 Mpc, our general strategy, particularly with larger error regions, appears less effective. However, increasing the exposure time to 250 s or 500 s may allow us to follow-up these triggers effectively. The increased exposure time will counteract the much fainter nature of these counterparts. This effect was not seen in Figure \ref{fig:measuredD}, likely due to that covering all error areas. In addition, the increased time to reach the correct field for areas greater than 1000 \degsq\,means the source is undetectable when it is reached, regardless of the exposure time.

Overall, therefore, we can significantly boost our chances of discovering an EM counterpart by being selective about which events we follow-up and, depending on the parameters of the trigger, by careful choice of exposure time. Selecting to prioritise sources with a high $P_{\rm DNS}$ will assist with this. This will also ensure \textit{Swift} remains available for other science.

\section{Discussion}
\label{sec:discussion}

While our method is thorough and replicates the real response to notices, there are a few points that should be addressed. These include whether and how \textit{Swift} should respond to NSBH GW triggers and what effect extinction might have on our results.

\subsection{NSBH mergers}

Significant EM emission is also expected from some NSBH merger in a similar way to a BNS merger. While in practice \textit{Swift}'s strategy is unlikely to differ when following-up an NSBH trigger versus a BNS trigger, we briefly examine possible differences here.

\textsc{MOSFiT} has recently been updated to include a semi-analytic forward NSBH merger model, \texttt{nsbh\_generative} \citep{Gompertz23}. Similarly to the \texttt{bns\_generative} model, light curves are predicted from the initial state of the binary and assumed equation of state of the NS. While we do not repeat our full analysis for this model, we can still draw significant conclusions on how \textit{Swift} will be able to respond to these triggers.

Like in a BNS merger, the ejecta in the NSBH merger can be divided into several components. These include tidal debris, similar to the BNS case, and two disk winds, one thermally driven and one magnetically driven. Both wind components have relatively low electron fractions with the thermal wind having a grey opacity and the magnetically driven wind being redder. A further significant difference between BNS and NSBH mergers is the impossibility of a post merger NS remnant in the latter case. As shown in Section \ref{sec:kilonova_model}, a significant proportion of BNS mergers will leave such a remnant which drives a neutrino wind to increase the ejecta's electron fraction and make the kilonova bluer. However, the effects of the magnetically driven wind and lack of possible remnant ensure the kilonova from a NSBH remains red with steady, little changing colours (see Figure 6 of \citealt{Gompertz23}).

Such a source is even more poorly suited to the UVOT filters than a BNS kilonova. This can be seen in Figure 5 of \citet{Gompertz23} which compares the peak magnitude of kilonovae from BNS and NSBH mergers. Even in the redder optical band presented there, NSBH kilonovae are much fainter by two to three magnitudes. This effect is likely to be greater for \textit{Swift}'s \textit{u} filter, for instance, and will mean many NSBH mergers will be too faint to detect. Thus, while we will respond to NSBH triggers during O4, we expect our chances of discovery to be extremely low. The strategy devised for BNS mergers is likely to be most effective as difficulty in merger classification means that our best chance is hoping the NSBH is a BNS in disguise.

\subsection{Binary black hole mergers}

To this point we have only considered the gravitational wave sources detectable by LVK that are expected to produce significant EM emission: BNS and NSBH mergers. However, the vast majority of sources detected by LVK are binary black hole (BBH) mergers. Rather than the kilonova and short GRB afterglow emission components examined here, the possible EM emission from a BBH merger is far more unconstrained. Due to the inherent nature of a BBH system, EM emission can only be produced if the merger occurs in a dense environment, for instance, the disk of an active galactic nucleus \cite[e.g.][]{Kelly17,McKernan19}. No confirmed EM counterpart to a BBH merger has thus far been discovered with only a few examples of possible candidates \citep[e.g.][see also \citealt{Greiner16} and \citealt{Connaughton18}]{Connaughton16,Graham20,Graham23,Cabrera24}. Due to the lack of constraints on these triggers' emission, \textit{Swift} follows a different strategy to that explored in this work. Instead \textit{Swift} follows-up only the best localised events with 90\% error areas $\leq 30$ \degsq and using minimal length 80 s exposures. This ensures the area is covered rapidly and minimises disruption to \textit{Swift}'s other activities. During O4, \textit{Swift} has followed up several BBH mergers meeting the criterion above, with no likely EM counterpart candidates so far.

\subsection{Extinction}

Another important aspect to consider is the reddening from dust and other interstellar and intergalactic media.  The wavelengths of the UVOT's filters mean that they will be affected to a large degree by such reddening and thus there will be a major impact on \textit{Swift}'s ability to detect an EM counterpart. We can minimise Galactic extinction by avoiding the Galactic plane where possible. This is already accounted for in the tiling algorithm which avoids the Galactic plane as much as possible. However, there is still the problem of extinction from the environment the merger took place in. This is impossible to predict as a wide variety of environments and extinctions have been observed in the short GRB population \citep[e.g.][]{Fong15}. Because the effect of this extinction cannot be eliminated, it is best to maintain our strategy without alterations. If another observatory were to identify an EM counterpart in a reddened environment, however, this would have to be accounted for in the targeted UVOT \citep[or XRT, see][]{Asquini19} follow-up that would occur.

\section{Conclusions}
\label{sec:conclusions}

The discovery of EM counterparts to GW events is vital to increase our understanding of both fundamental physics and the impact these sources have on the universe, for instance, in terms of elemental abundances.

\textit{Swift} has distinct advantages offered by its nature as a satellite. It does not need to wait for night and is, at present, the only way to catch the early blue and UV emission. Its rapid response offers a window unprobed by other observatories. However, the UVOT's relatively narrow field of view means often thousands of fields will need to be observed and the red nature of the predicted source may make it hard to discern. \textit{Swift} is vitally important for many other fields of X-ray and UV astronomy and it is important to make use of it effectively.

We have therefore shown that we can maximise \textit{Swift}'s efficiency in discovering counterparts to GW triggers by careful target selection. In particular, targeting only triggers with inferred distances $\leq300$ Mpc and a sky localisation $\leq500$ \degsq\,should result in $\sim25$\% of counterparts being discovered. Extending this to greater distances requires cutting the area to $<150$ \degsq\,which still reduces the proportion that may be detected significantly. For triggers with a low $P_{\rm DNS}$, limiting our follow-up to only the closest and therefore brightest sources is also necessary. We further find the optimal exposure time to be 120 s, although this is dependent on the properties of the trigger, and that the \textit{u} filter offers the best balance between source brightness and background. We note that strictly obeying these criteria does restrict follow-up to a minority of triggers. However, as detector sensitivity improves and particularly as Virgo and KAGRA begin observations in earnest, sky localisation will improve dramatically and the fraction of triggers to which our criteria apply will significantly increase. This will continue as LVK approach their final sensitivity in future observing runs and while many of our conclusions will still apply, it will be important to revisit how they are affected by these changes.

In addition to improvements to the gravitational wave detector network, the launch of other missions, particularly UV focussed, will have a significant impact on how GW events are followed up. Our results show that even a narrow field instrument like the UVOT can effectively follow-up GW triggers. This bodes extremely well for the discovery potential of sensitive, wide-field missions such as the \textit{Ultraviolet Explorer} \citep[\textit{UVEX},][]{Kulkarni21} or the \textit{Ultraviolet Transient Astronomy Satellite} \citep[\textit{ULTRASAT,}][]{Shvartzvald24}. Even with these missions operational, however, \textit{Swift}'s UVOT will still be crucial for characterising the spectral and temporal evolution of kilonovae, particularly at early times. Following up the most likely candidates from such missions may therefore be the best strategy in future LVK observing runs.

While to date O4 has been somewhat disappointing in terms of the triggers the universe has supplied, there is still a significant amount of time remaining and new EM bright events could occur at any time. In this work, we have ensured \textit{Swift} is in an optimal position to respond to future events.

\section*{Acknowledgements}

This work made use of data supplied by the UK \textit{Swift} Science Data Centre at the University of Leicester; and the Wide Field Astronomy Unit (WFAU).

RAJEF acknowledges support from the UK Space Agency and the European Union’s Horizon 2020 Programme under the AHEAD2020 project (grant agreement number 871158). ET acknowledges support by the European Research Council through the Consolidator grant BHianca (Grant agreement ID: 101002761).

\section*{Data Availability}
The simulated compact binary mergers are available at \url{https://zenodo.org/record/7026209#.Y90gIKfP1hE}.

The Two Micron All Sky Survey Photometric Redshift catalogue (2MPZ) is available at \url{http://ssa.roe.ac.uk/TWOMPZ.html}.



\bibliographystyle{mnras}
\bibliography{main}



\appendix



\bsp	
\label{lastpage}
\end{document}